\documentclass[twocolumn,secnumarabic,amssymb, nobibnotes, aps, prc]{revtex4-1}
\usepackage{graphicx}
\usepackage{lineno}

\begin{document}

\title{Two-neutrino double-beta decay of $^{150}$Nd to excited final states in $^{150}$Sm}


\author{M.F. Kidd}
\email{Now at Tennessee Technological University, PO Box 5051, Cookeville, TN 38505}
\email{mkidd@tntech.edu}
\author{J.H. Esterline}
\author{S. W. Finch}
\author{W. Tornow}
\address{Department of Physics, Duke University, Durham, North Carolina 27708, USA}
\address{Triangle Universities Nuclear Laboratory, Durham, North Carolina 27708, USA}

\date{\today}

\begin{abstract}
\begin{description}
\item[Background]
Double-beta decay is a rare nuclear process in which two neutrons in the nucleus are converted to two protons with the emission of two electrons and two electron anti-neutrinos.  
\item[Purpose]
We measured the half life of the two-neutrino double-beta decay of $^{150}$Nd to excited final states of $^{150}$Sm by detecting the de-excitation gamma rays of the daughter nucleus.  
\item[Method]
This study yields the first detection of the coincidence gamma rays from the 0$^+_1$ excited state of $^{150}$Sm.  These gamma rays have energies of 333.97 keV and 406.52 keV, and are emitted in coincidence through a 0$^+_1\rightarrow$2$^+_1\rightarrow$0$^+_{gs}$ transition.
\item[Results]
  The enriched Nd$_2$O$_3$ sample consisted of 40.13 g $^{150}$Nd and was observed for 642.8 days at the Kimballton Underground Research Facility, producing 21.6 net events in the region of interest.  This count rate gives a half life of $T_{1/2}=(1.07^{+0.45}_{-0.25}(stat)\pm0.07(syst.))\times 10^{20}$ years.  The effective nuclear matrix element was found to be 0.0465$^{+0.0098}_{-0.0054}$.  Finally, lower limits were obtained for decays to higher excited final states. 
\item[Conclusions]
Our half-life measurement agrees within uncertainties with another recent measurement in which no coincidence was employed.  Our nuclear matrix element calculation may have an impact on a recent neutrinoless double-beta decay nuclear matrix element calculation which implies the decay to the first excited state in $^{150}$Sm is favored over that to the ground state.  

\end{description}
\end{abstract}


\maketitle


\section{Introduction}\label{sec:intro}

The main motivation for studying double-beta decay is clear:  to shed light upon the nature of the neutrino.  Though much has been discovered in the field of the neutrino since its conception and later discovery, some very basic traits remain unknown.  Two of these are the Majorana or Dirac nature of the neutrino, and the particle's mass.  Observation of neutrinoless double-beta (0$\nu\beta\beta$) decay would answer the question of the nature of the neutrino while simultaneously determining the mass, assuming the nuclear matrix elements (NMEs) for that particular nuclear transition are known.  Here, the study of two-neutrino double-beta (2$\nu\beta\beta$) decay can prove useful by providing experimental data which can be used to test and calibrate theoretical models needed to calculate 0$\nu\beta\beta$ NMEs \cite{Rod03}.  A study of the rate of the 2$\nu\beta\beta$ decay of $^{150}$Nd to excited final states of $^{150}$Sm will be described.  This work took place both at Triangle Universities Nuclear Laboratory (TUNL) and at Kimballton Underground Research Facility (KURF).  

The 2$\nu\beta\beta$ decay rate, $\lambda$, can be described by the following equation,
\begin{equation}
\label{equ:2nbbdecayratePS}
\lambda=G^{2\nu}|M^{2\nu}_F-M^{2\nu}_{GT}|^2,
\end{equation}
where $G^{2\nu}$ contains phase space integrals and relevant constants and is dependent on the Q-value of the decay, and $M^{2\nu}_F-M^{2\nu}_{GT}$ represents the Fermi (F) and Gamow-Teller (GT) components of the 2$\nu\beta\beta$ decay NMEs.  Ideal nuclei for 2$\nu\beta\beta$ decay studies have a high Q-value and large NMEs.  There are 35 nuclei that undergo 2$\nu\beta\beta$, but only 11 ($^{48}$Ca, $^{76}$Ge, $^{82}$Se, $^{96}$Zr, $^{100}$Mo, $^{110}$Pd, $^{116}$Cd, $^{124}$Sn, $^{130}$Te, $^{136}$Xe, and $^{150}$Nd) have a Q-value which is practical for experimental use.  This Q-value restriction is about 2 MeV; below that, such a rare decay could get overwhelmed by natural radiation.  Of the 11 aforementioned nuclei, the double-beta decay to the ground state of ten has been measured:  $^{48}$Ca, $^{76}$Ge, $^{82}$Se, $^{96}$Zr, $^{100}$Mo, $^{116}$Cd, $^{128}$Te, $^{130}$Te, $^{150}$Nd, $^{136}$Xe and (not previously listed) $^{238}$U (see \cite{Bar10} for an excellent compilation of these results).  However, the neutrinoless mode has not yet been observed.  

Besides the Q-value, other quantities which can be considered in order to maximize the experiment's success in detecting double-beta decay include natural abundance, availability of the isotope, possibility of enrichment, and source cost.  To optimize the source efficiency, the number of isotope nuclei must be maximized, which is heavily dependent on the abundance, enrichment possibility, and availability of the isotope.  One of the best candidates for 0$\nu\beta\beta$ searches is $^{150}$Nd.  It has the second highest Q-value (Q$_{\beta\beta}=$3.37 MeV) and the largest phase-space factor of all 0$\nu\beta\beta$ candidates \cite{Fan11}, and has a natural abundance of 5.64\%.  Although $^{150}$Nd is strongly deformed, calculations by Fang {\it et al.} \cite{Fan11} show that the deformation suppresses the 0$\nu\beta\beta$ NME by only about 40\%.  

Very recently, a novel decay channel of the scissors mode 1$^+_{sc}$ to the first excited 0$^+$ state (0$^+_1$) of $^{154}$Gd was reported by Beller {\it et al.} \cite{Bel13}, implying a much larger matrix element than previously thought for the 0$\nu\beta\beta$ decay to the 0$^+_1$ state in the shape transitional regions of the N=90 isotones.  In fact, it was argued in \cite{Bel13} that the 0$\nu\beta\beta$ decay of  $^{150}$Nd to the 0$^+_1$ state in $^{150}$Sm may be slightly favored over the transition to the 0$^+$ ground state due to the calculated nuclear matrix element ratio M$^{0\nu}(0^+_1)/M^{0\nu}(0^+_{gs}$)=1.6.  If the reduced phase space for the 0$^+_1$ transition is taken into account, the ratio of the partial decay rates can be calculated using $\lambda_{0\nu\beta\beta}[0^+_1]/\lambda_{0\nu\beta\beta}[0^+_{gs}]=(G_{0\nu}(0^+_1)|M^{(0\nu)}[0^+_1]|^2)/(G_{0\nu}(0^+_{gs})|M^{(0\nu)}[0^+_{gs}]|^2)=1.2$ \cite{Bel13}.  Measurements of the 2$\nu\beta\beta$ decay of $^{150}$Nd to the 0$^+_1$ state in $^{150}$Sm may shed light on this conjecture.  

\begin{figure}[h]
\centering
\includegraphics
[width=3.25in]{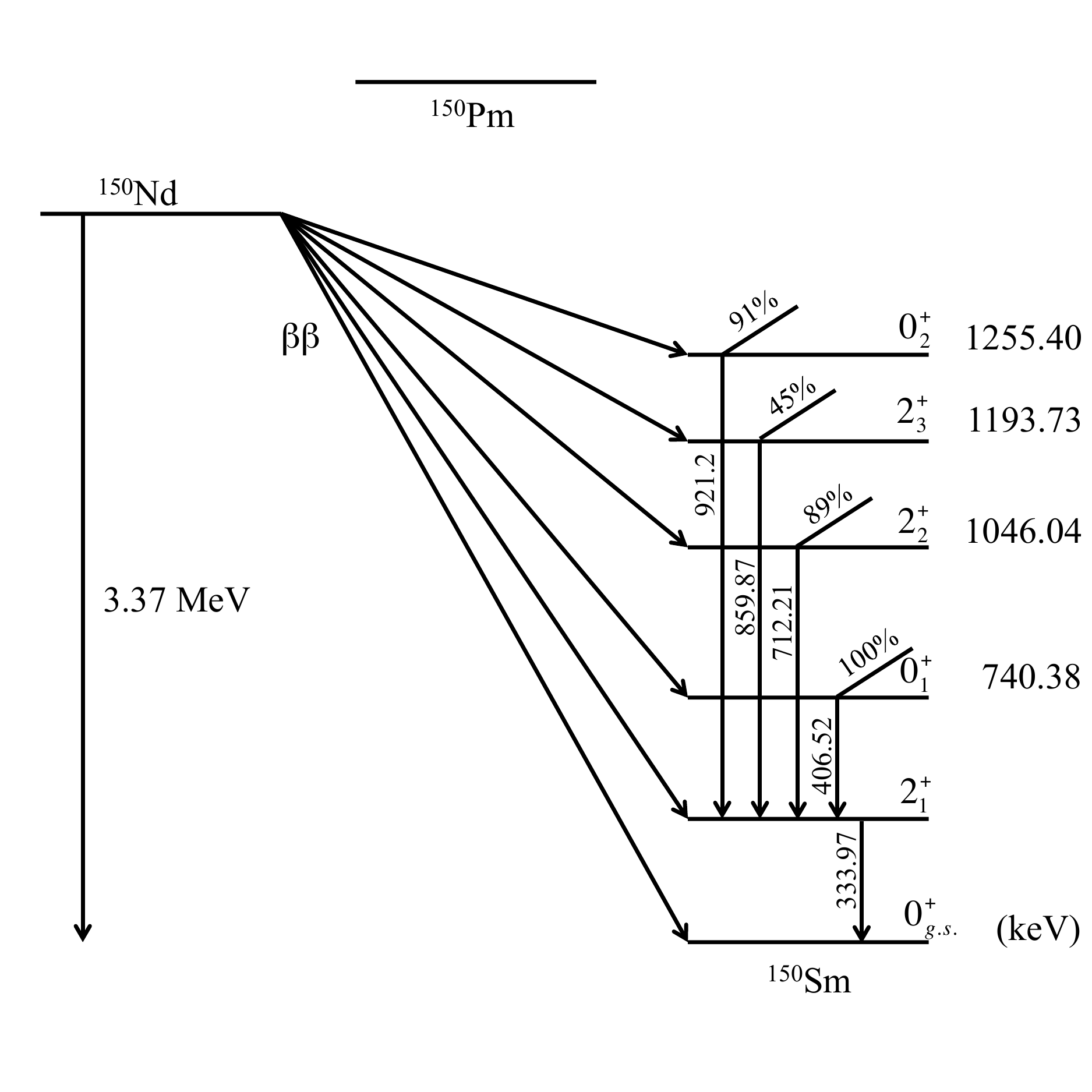}
\caption[Higher excited states of $^{150}$Nd.]{Level scheme of $^{150}$Nd double-beta decay to excited states of $^{150}$Sm.  }
\label{fig:nd-higher}
\end{figure}

Recently, the 2$\nu\beta\beta$ decay half life of $^{150}$Nd to the excited 0$^+_1$ state of $^{150}$Sm has been measured by Barabash {\it et al.} \cite{Bar09} to be T$_{1/2}=(1.33^{+0.36}_{-0.23}(stat)^{+0.27}_{-0.13}(syst))\times 10^{20}$ y.  Limits have been established for 2$\nu\beta\beta$ decay to other excited final states of $^{150}$Sm.  (see Fig. \ref{fig:nd-higher} and Table \ref{table:previous_Nd150})  Previously, only limits had been reported for the decay to the 0$^+_1$ state of $^{150}$Sm \cite{Kli02},\cite{Isa96},\cite{Arp94},\cite{Arp99}.  

\begin{table}[t]
\caption[Previously measured values for $^{150}$Nd 2$\nu\beta\beta$ half lives.]{Previous values for $^{150}$Nd 2$\nu\beta\beta$ half lives as measured by NEMO \cite{NEMO08} and Barabash {\it et al.}  \cite{Bar09}}
	\begin{center}
	\begin{tabular}{c|c|c}
			\hline
			Final  & $\gamma$-ray Energy & $(T^{0\nu+2\nu}_{1/2})_{exp}$ (y) \\
                        State&  (keV) & previous works \\
                          in $^{150}$Sm&&\\\hline
			0$_{g.s.}$    & 0.0                & $(9.11^{+0.25}_{-0.22}(stat)\pm0.63(syst))\times10^{18}$\cite{NEMO08}\\
			2$^+_1$       & 333.97             & $> 2.2\times10^{20}$\cite{Bar09}\\
			0$^+_1$       & 333.97, 406.52      & $(1.33^{+0.36}_{-0.23}(stat)^{+0.27}_{-0.13}(syst))\times 10^{20}$\cite{Bar09}\\
			2$^+_2$       & 712.21, 333.97     & $>8.0\times10^{20}$\cite{Bar09}\\
                        2$^+_3$       & 1193.83            & $>5.4\times10^{20}$\cite{Bar09}\\
                        0$^+_2$       & 921.2, 333.97      & $>4.7\times10^{20}$\cite{Bar09}\\
		\end{tabular}
	\end{center}
	
	\label{table:previous_Nd150}
\end{table}

\section{Experimental Method}\label{sec:method}

Our double-beta decay setup consists of two high-purity germanium (HPGe) detectors surrounded by several layers of active and passive shielding.  These detectors operate in coincidence in order to detect de-excitation $\gamma$ rays from an excited final state of a nucleus which has undergone double-beta decay.  The implementation of the  coincidence technique here has two results.  Requiring two distinct, simultaneous gamma-ray energy deposits immediately and significantly reduces the occurrence of background events.  Also, although the efficiency for detecting these two gamma rays is reduced compared to single-gamma detection, the uniqueness of the signal means a generally unambiguous result.  

\begin{figure}
\centering
\includegraphics
[width=3.5in]{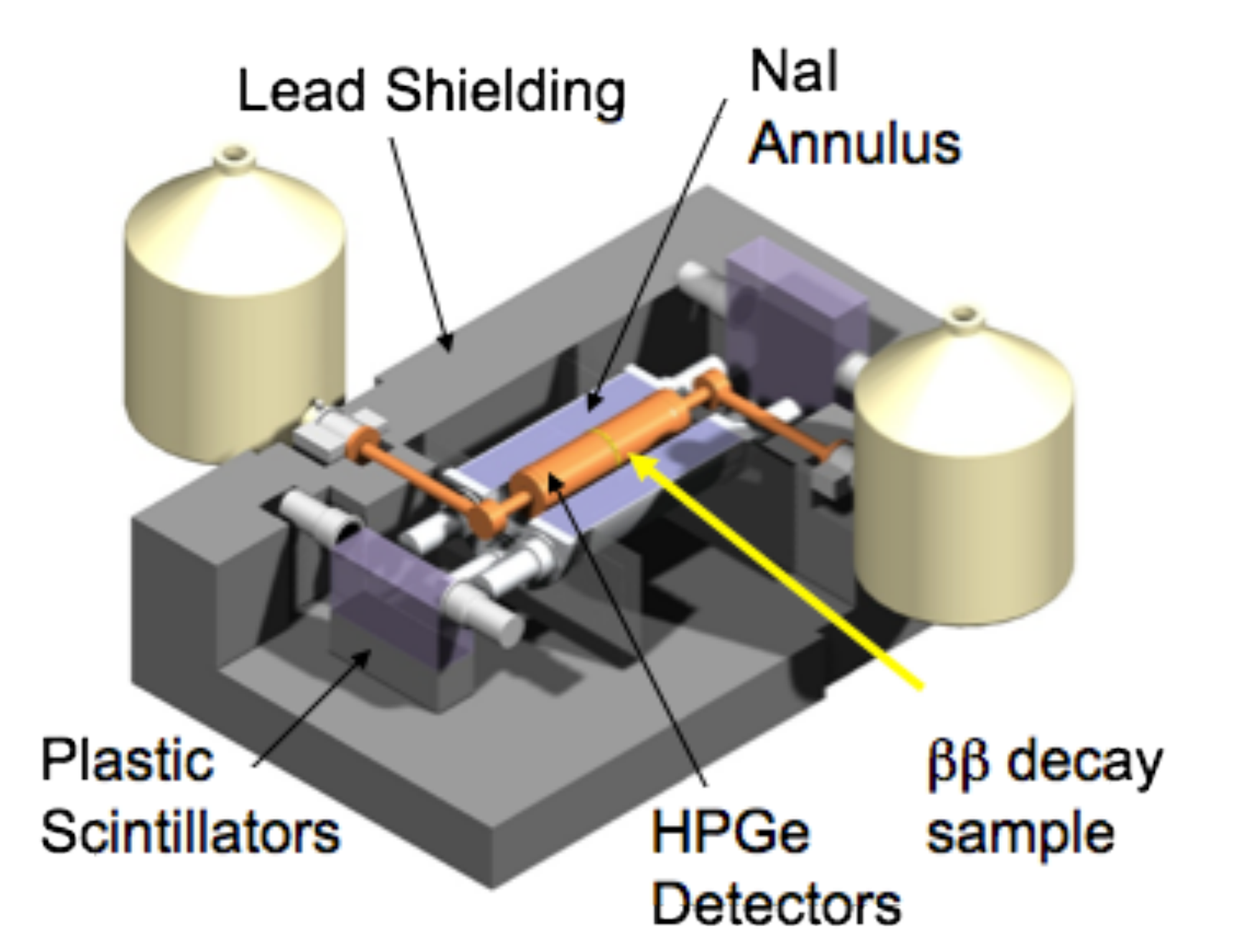}
\caption[Diagram of setup.]{(Color online.)  Diagram of present double-beta decay setup.  Not pictured are the OFHC copper plates between the NaI annulus and the lead shield.  (Not to scale.  Please refer to text for dimensions.)}
\label{fig:setup}
\end{figure}

The HPGe detectors of the TUNL double-beta decay setup are p-type, coaxial germanium detectors about 8.8 cm in diameter and 5.0 cm in thickness.  The endcap of each detector is nickel-plated magnesium about 2.54 mm thick at the front face.  The detector assembly is connected to the cryostat in the j-type configuration.  To reduce background, the preamplifier and high-voltage filter are housed away from the crystal near the LN dewar.  This type of configuration also easily allows the detectors to be inserted into the veto annulus.  A diagram of the setup is shown in Figure \ref{fig:setup}.

\begin{figure}
\centering
\includegraphics
[width=3.5in]{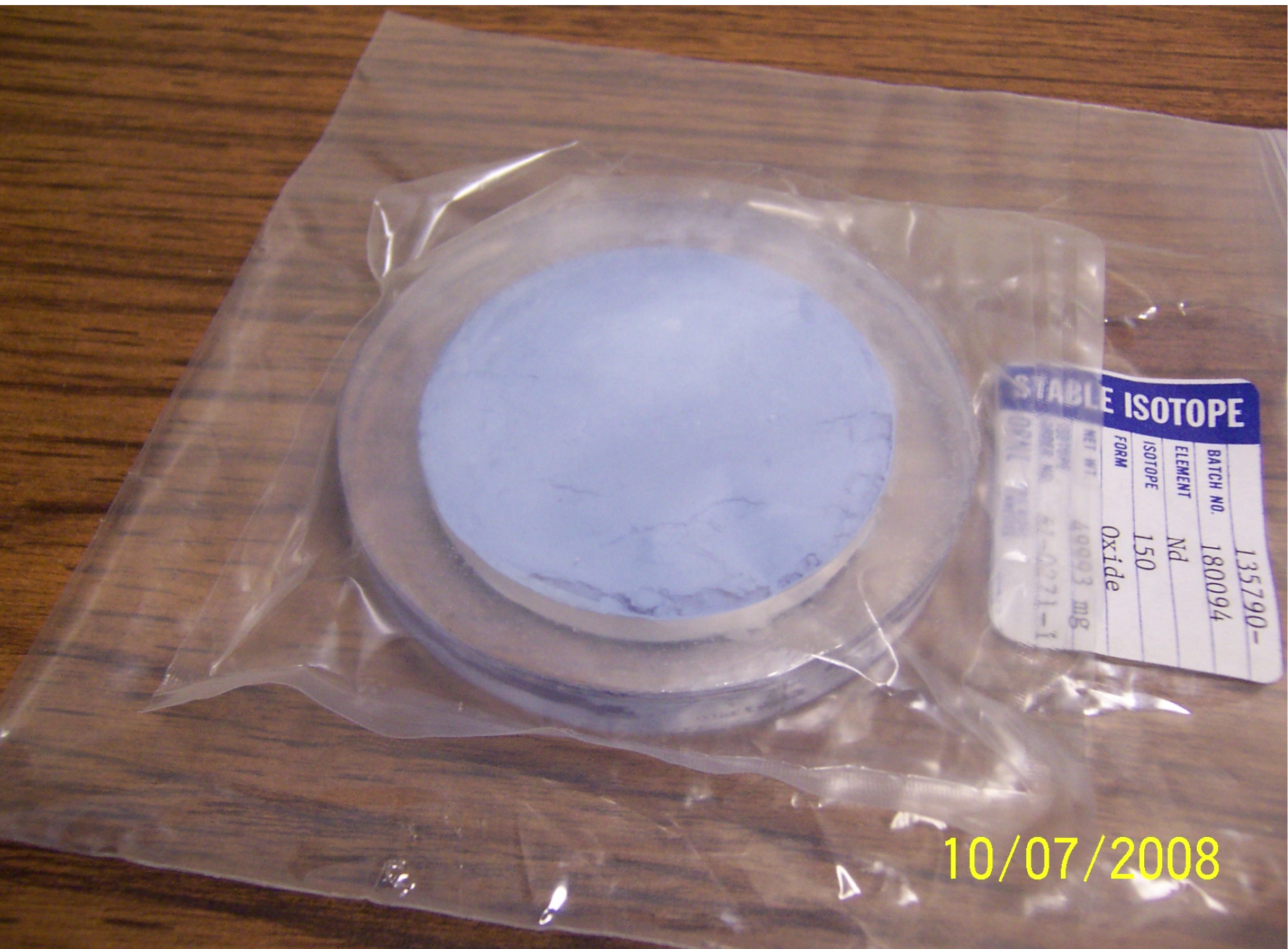}
\caption[150Nd Sample]{(Color online.)  Photograph of Nd$_2$O$_3$ sample (bluish powder) encased in polycarbonate and tedlar.}
\label{fig:sample}
\end{figure}

The detectors sandwich the double-beta decay sample.  Our neodymium sample consists of 50.00 g of Nd$_2$O$_3$ powder, which corresponds to 42.87 g of Nd.  This is enriched to 93.60\% $^{150}$Nd, which is 40.13 g $^{150}$Nd.  The Nd$_2$O$_3$ powder is compressed within a cylindrical cavity in a polycarbonate holder 0.780 $\pm$ 0.005 cm thick and 2.86 cm in radius.  The polycarbonate casing adds 0.159 $\pm$ 0.005 cm to each side.  Though the total diameter of the holder is about 8 cm, the Nd$_2$O$_3$ is concentrated in the center to take advantage of the coincidence efficiency of the detectors (see Sec. \ref{sec:coineff}). Furthermore, to guard against losing any of the enriched Nd$_2$O$_3$ if the polycarbonate was breached, the sample was kept within a tedlar bag, which added 0.013 $\pm$ 0.005 cm to each side.  See Fig. \ref{fig:sample} for a picture of the Nd$_2$O$_3$ sample.  The detectors were then 1.124 $\pm$ 0.009 cm apart.

\subsection{Passive shielding}
\label{sec:pass}
The first goal of shielding is to reduce the number of gamma rays produced outside the sample from reaching the detectors.  To absorb these gamma rays, high-Z material is placed around the detector setup.  The TUNL setup has a lead house built around and beneath it.  The thickness is about 6 inches (about 15 cm) on all sides.  At 500 keV, relatively near our region of interest, the attenuation of 15 cm of lead is $>$10$^6$.  A likely contaminant in lead shielding is $^{210}$Pb, which has a half-life of 22.2 years.  The decay chain of $^{210}$Pb includes some high-energy beta decays, which can result in Bremsstrahlung radiation.

Within the lead shield, there is a layer of 3/4 inch thick Oxygen Free High Conductivity (OFHC) copper plates.   OFHC copper has a purity of 99.99\%, but can still be activated via (n,$\alpha$) to $^{60}$Co by cosmic-ray neutrons.  Aside from this disadvantage, OFHC copper has low concentrations of $^{208}$Tl ($<$0.005 Bq/kg), $^{214}$Bi ($<$0.02-0.17 Bq/kg), and $^{40}$K ($<$0.2 Bq/kg)\cite{Gil08}.  Using another absorbing material, such as copper inside the lead shield can reduce the effect of the Bremsstrahlung radiation from the daughters of $^{210}$Pb.  

\subsection{Active shielding}
\label{sec:active}

Another way to reduce the background, especially the Compton continuum, is to employ active shielding.  This type of shielding can discriminate when a gamma ray is scattered out of the detector leaving only part of the energy deposited, or when a gamma ray comes from outside the system.  The HPGe detectors are surrounded by a sodium iodide (NaI) annulus with two plastic plate scintillators on the endcaps.  The NaI(Tl) crystal is housed inside of low-background aluminum.  The annulus has dimensions of 12.5 cm and 35.6 cm for the inner and outer diameters, respectively, and is 50 cm long.  Six standard 3 inch diameter photomultiplier tubes (PMTs) are installed on each end of the NaI annulus, though using only three on each side provided adequate coverage.  The plastic plate scintillators are 10 cm thick and 30 cm $\times$ 30 cm square.  Each plastic plate has two 2 inch PMTs on opposite sides.  

Because of its proximity to the HPGe detectors, the NaI annulus is particularly vital to the veto process.  The plastic scintillators function to veto particles which would not be seen by the annulus, {\it i.e.}, they travel nearly horizontal, along the axis of the annulus.  This seems unlikely for a primary particle, but secondary interactions in the shielding, or Compton scattering in the detectors could result in such a condition.  

\subsection{Electronics}\label{sec:electronics}

Our electronics setup consists of standard commercially available NIM and Computer Automated Measurement and Control (CAMAC) modules.  A signal produced in one of the HPGe detectors is first transmitted through a preamplifier and is then sent to an amplifier for shaping and amplification.  There are two outputs of the amplifier module.  The unipolar output is sent to the ADC via a delay amplifier set to 4.75 $\mu$s.  This delay is chosen to coordinate the signal with the separate timing signal which opens the gate to the ADC.  The pulse height of this unipolar signal is proportional to the energy deposited in the HPGe detector.  

\begin{figure*}
\centering
\includegraphics
[height=8.5in]{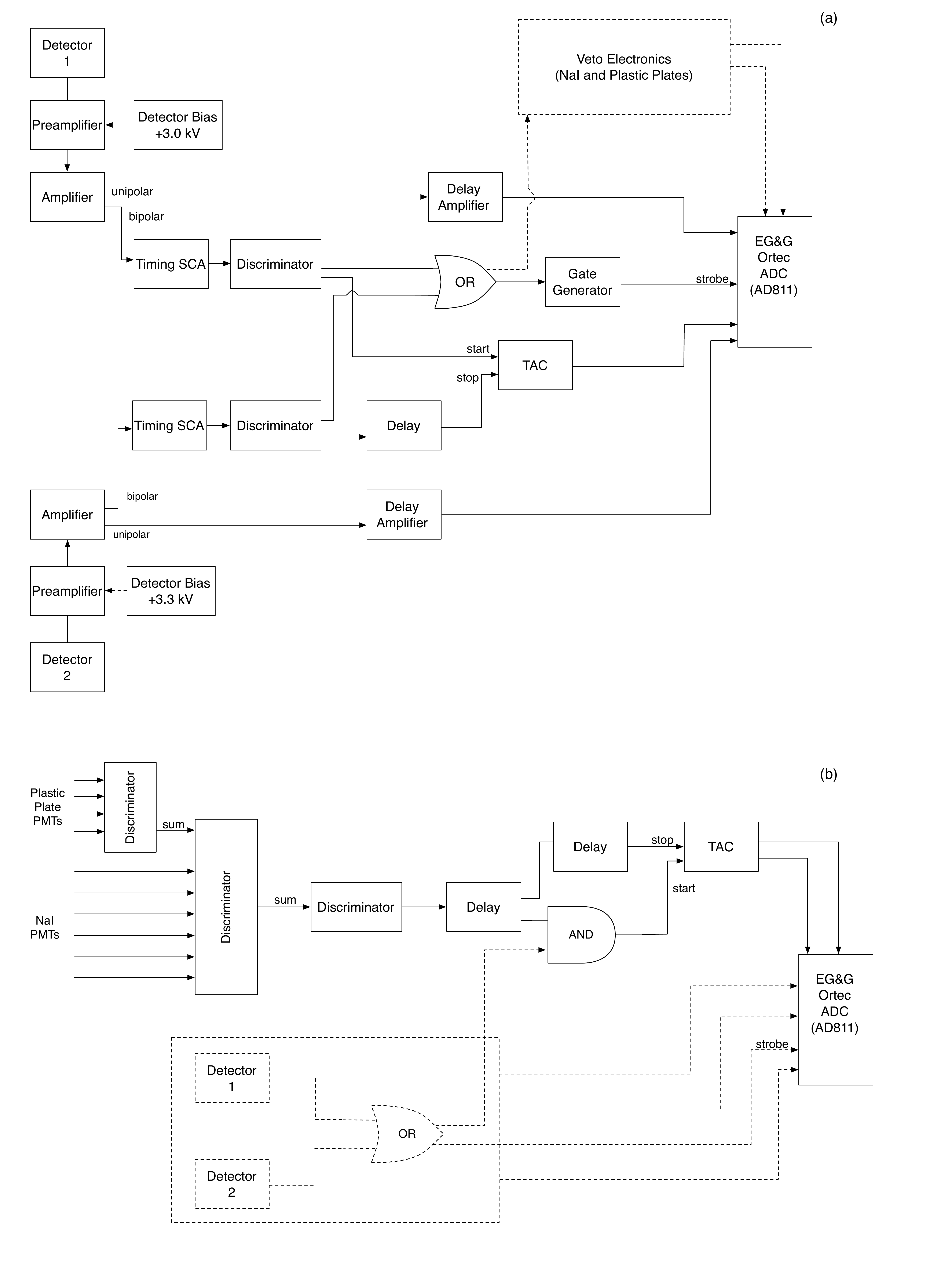}
\caption[Electronics]{(a) Diagram showing the electronics for the primary coincidence between the HPGe detectors.  (b) Diagram showing the electronics for the veto detectors surrounding the HPGe detectors.  See Sec. \ref{sec:electronics} for details.  }
\label{fig:electronics}
\end{figure*}

The bipolar output of the amplifier is sent first to a timing single-channel analyzer (SCA) whose output is a logic signal which represents the arrival of the original signal.  This signal is sent to a discriminator.  One of the discriminator outputs is sent to start the time-to-amplitude converter (TAC).  For detector 1, this signal is sent directly to the ``start" input of the TAC, but for detector 2, the corresponding signal is first directed through a delay of 1 $\mu$s before being sent to the ``stop" of the TAC.  The TAC output is sent to the ADC.  The other discriminator output goes to a logic module which is set to deliver a logic pulse if there is a signal in either HPGe detectors.  One output of this OR gate triggers the ADC, and the other goes to another logic circuit which produces a logic signal if there is also a corresponding signal in the veto electronics.  (See Fig. \ref{fig:electronics} a.)

The intention of this coincidence measurement is to detect a signal in coincidence with both HPGe detectors, but not in any of the veto detectors.  Therefore, the energy of the signal coming from the veto detectors is not relevant; only the fact that the veto counters have fired in coincidence with either HPGe detector is important.  The outputs of the PMTs which collect the scintillation light from both the NaI crystal and the plastic shields are summed, sent to a discriminator, and delayed to coincide with the signals from the HPGe detectors.  This output is sent to the logic unit mentioned in the previous paragraph which performs an AND gate on the veto and HPGe signals.  If there is a signal in either HPGe detector and a signal in the veto counters, an output is sent to start the veto TAC.  The stop is a delayed copy of the veto signal, and the TAC output is sent to the ADC.  (See Fig. \ref{fig:electronics} b.)

\subsection{Computer interface}

After making it to the ADC, the signal must now be digitized and sent to the data acquisition (DAQ) system.  The ADC used in this setup is an Ortec AD811.  This module is a CAMAC module in a CAMAC crate.  It contains eight ADCs which can each accommodate 11 bits, or 2047 channels.  Of these eight ADCs, five are utilized:  one for each HPGe detector, one for the timing between the two HPGe detectors, and two for the timing between either HPGe and the veto counters.  

The strobe input is supplied by the gate generator in the electronics circuit which is triggered by a signal from either HPGe detector.  The strobe input begins the digitization process of any peaks found in the five ADC channels.  The AD811 takes approximately 80 $\mu$s to process an analog signal and output a digital signal, and afterwards generates a ``look-at-me" (LAM) signal to the CAMAC controller.  The CAMAC crate controller is a Wiener CC32.  The CC32 module interfaces the CAMAC crate with the data-acquisition host computer.  

The DAQ system used is the CEBAF On-line Data Acquisition (CODA) program, first developed at Jefferson Laboratory in Newport News, Virginia.  Each CODA data file contains the energy deposited in each detector and the three TAC spectra.  These files are converted to ROOT files for analysis via the TUNL Real-time Analysis Package (TRAP).  TRAP creates a ROOT tree which contains event-by-event data for each detector \cite{TRAP}.

\subsection{Kimballton Underground Research Facility}\label{KURF}

The Kimballton Underground Research Facility (KURF) is a relatively new facility located at the Kimballton mine run by Lhoist North America.  The Kimballton mine is an operating limestone mine with over 50 miles of drifts and a current maximum depth of 2300 feet.  Our present facility is located at the 14th level at a depth of 1700 feet [1450 meters water equivalent (m.w.e.)].  For more information about the facility, please consult Ref. \cite{Finnerty201165}.


\begin{figure*}
\centering
\includegraphics
[width=7in]{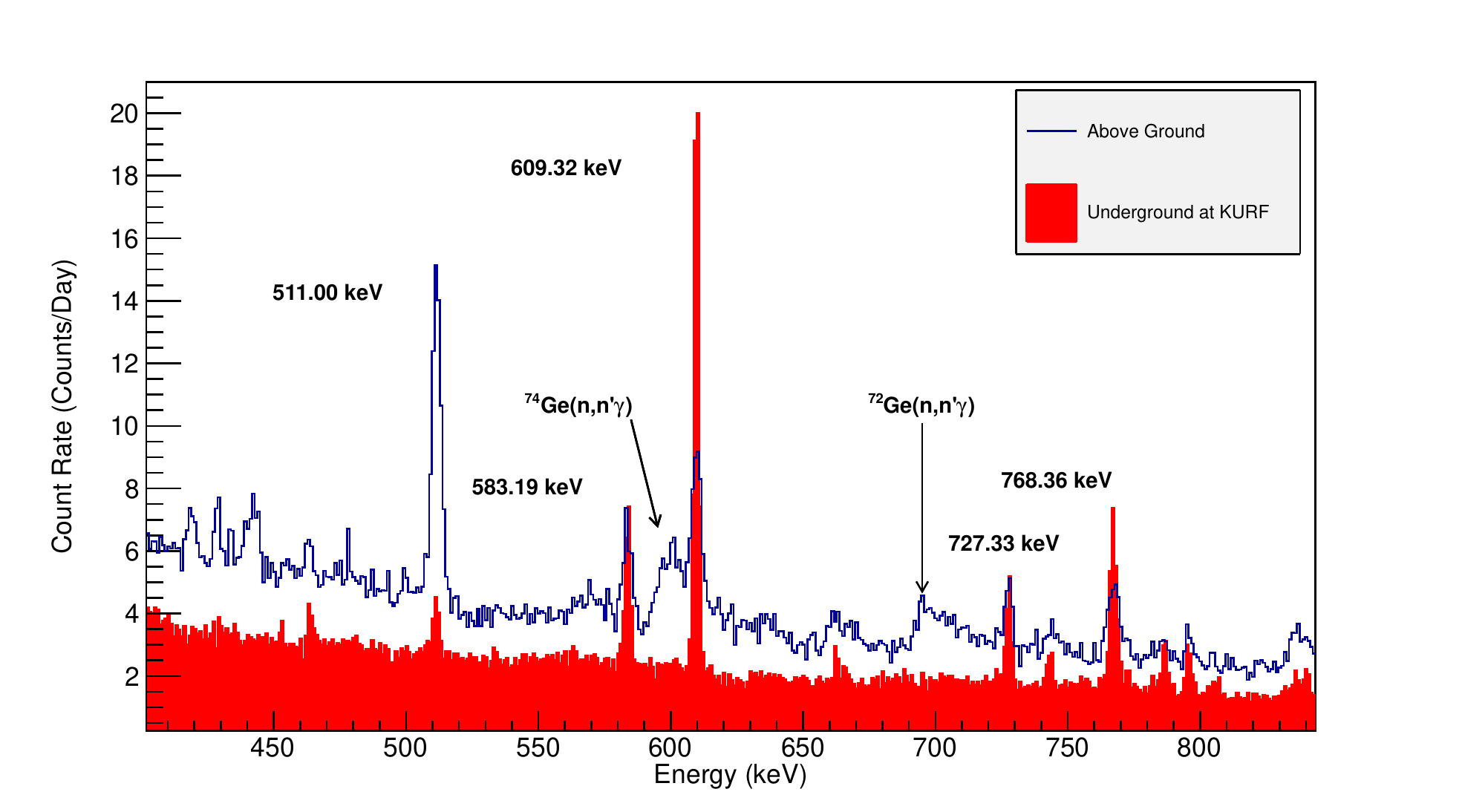}
\caption[A comparison of data taken at ground level and at Kimballton Underground Research Facility]{(Color online)  The reduction of the 511 keV $\gamma$--ray line and the asymmetric peak from inelastic neutron scattering on $^{74}$Ge and $^{72}$Ge is evident.  The labeled peaks in the spectrum are identified as electron-positron annihilation (511.00 keV), the beta decay of $^{208}$Tl to ${208}$Pb (583.19 keV), the beta decay of $^{214}$Bi to $^{214}$Po, (609.32 keV, 768.36 keV), and the beta decay of $^{212}$Bi to $^{212}$Po (727.33 keV).}
\label{fig:zoom}
\end{figure*}

Our double-beta decay apparatus was previously operated above ground in the TUNL Low Background Counting Facility (LBCF), a shielded room in the basement of the Duke Physics Building where it was used to investigate the double-beta decay of $^{100}$Mo to excited final states  \cite{Kid09} and neutrinoless double-electron capture in $^{112}$Sn \cite{Kid08}.  Fig. \ref{fig:zoom} shows normalized singles spectra taken at the TUNL LBCF and at KURF with an enriched $^{112}$Sn sample in place.  While many of the gamma--ray lines in Fig. \ref{fig:zoom} are intrinsic to our setup and thus not reduced by moving underground, the reduction of cosmic--ray background is apparent.  The 511.00 keV peak is reduced by a factor of ten, and the inelastic neutron scattering by $^{74}$Ge at 596.85 keV and $^{72}$Ge at 689.6 keV is no longer distinct as can be seen in Fig. \ref{fig:zoom}.  However, the radon concentration (see 609.32 keV line from the decay of $^{214}$Bi to $^{214}$Po in the $^{238}$U decay chain) at KURF is considerably larger than above ground at TUNL.

\subsection{Analysis}\label{sec:analysis}

Data acquisition took the form of runs of 3-5 days in length.  A run over this length of time resulted in the accumulation of sufficient counts for calibration purposes.  In the TUNL setup, there are a few contaminants that are inherent to the setup or to the sample that are used for calibration purposes.  A wide range of energies were chosen to ensure a calibration valid over the entire spectrum.  Peaks used were the 238.63 keV from the $\beta$-decay of $^{212}$Pb in the $^{232}$Th decay chain, the annihilation peak at 511.00 keV, the $^{40}$K peak at 1460.82 keV, and the peak at 1764.49 keV from the $\beta$-decay of $^{214}$Bi to $^{214}$Pb.  In addition to providing calibration data, such frequent calibration allowed for monitoring of detector stability.  Each calibration peak was fit with a Gaussian with a linear background.  The centroids extracted from the fit of each calibration peak as well as the calibration fit parameters were stored for each run.  The 238.63 keV peak's centroid varied by only 0.2\% over the 642.8 day counting period, and the 511.00 keV peak varied by only about 0.3\%.  

\begin{figure}
\centering
\includegraphics
[width=3.5in]{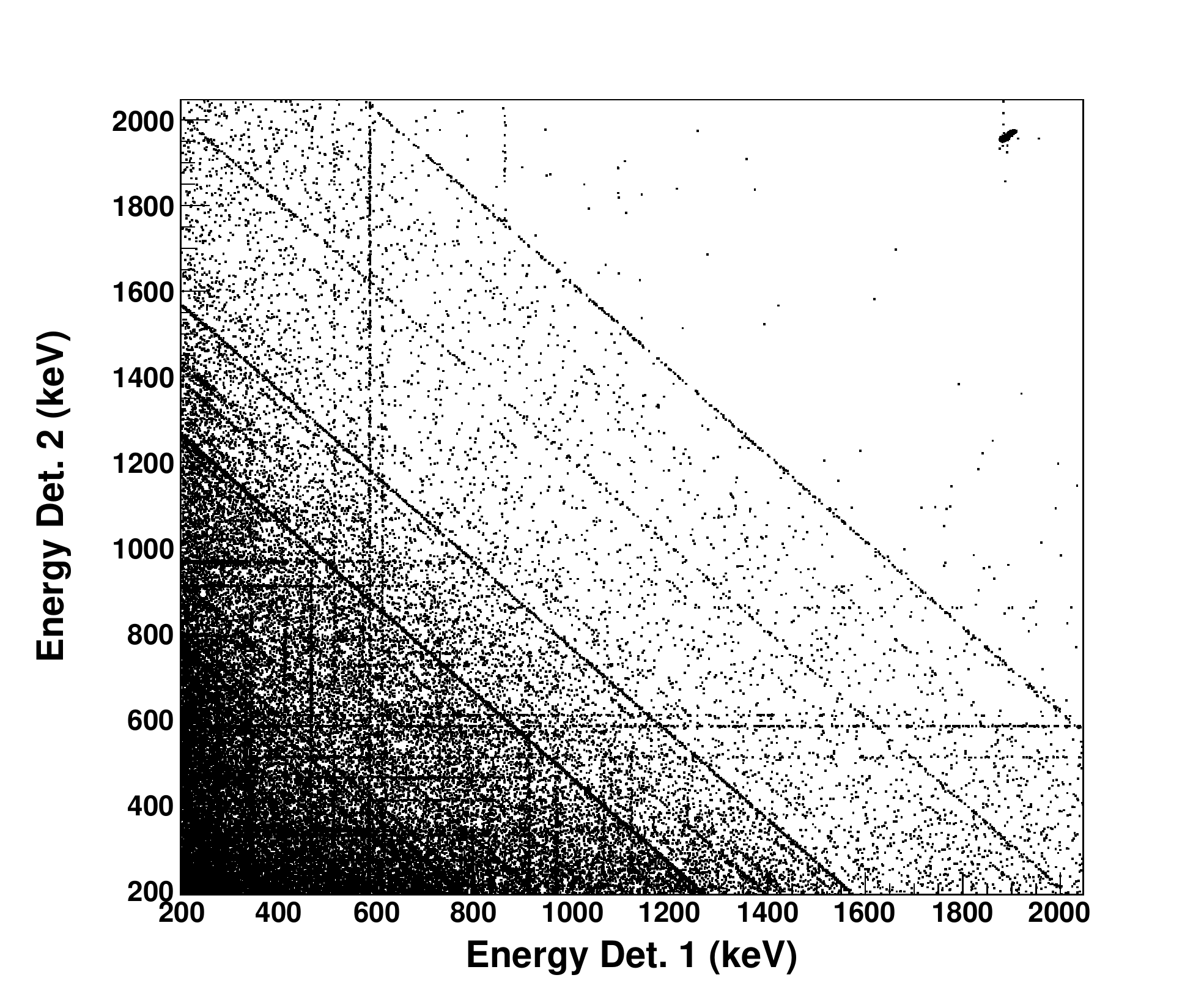}
\caption[An example of a two-dimensional spectrum.]{An example of a two-dimensional spectrum taken at KURF
 for about 250 days.  Energy in detector 2 is plotted against energy in detector 1.  Bins are 1 keV x 1 keV.  See text for an explanation of the features of the plot.  }
\label{fig:2Dspectrum}
\end{figure}
\begin{figure*}
\centering
\includegraphics
[width=7in]{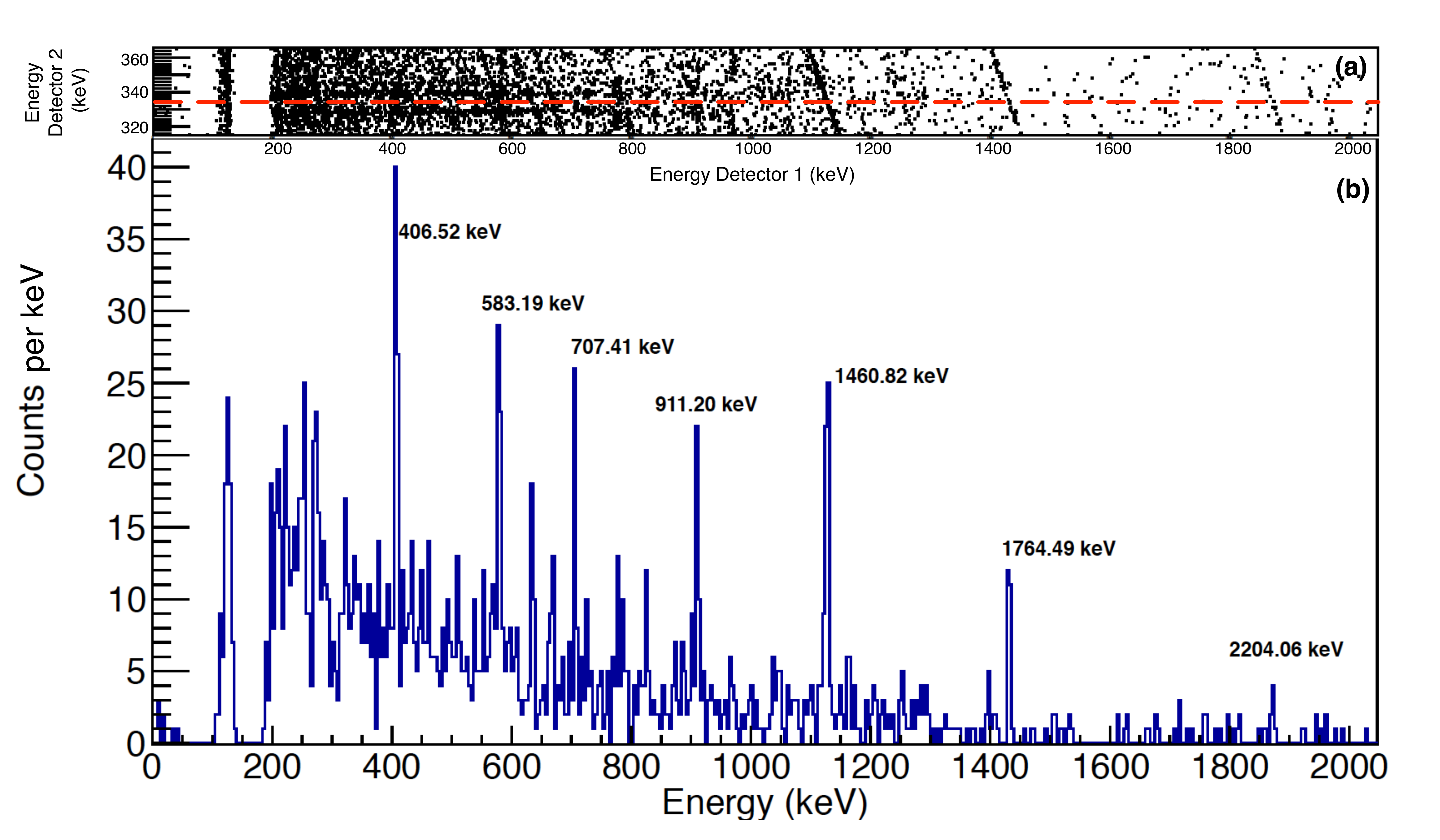}
\caption[Example]{(Color online) (a) A slice of the two-dimensional spectrum of Fig. \ref{fig:2Dspectrum}.  The dashed line shows 333.97 keV.  The resulting projected full-energy spectrum in coincidence with 333.97 keV is shown in (b).  Here the energy scale is 4 keV per bin.  Here it can be seen that a cross section of the two-dimensional spectrum containing the diagonal and vertical lines can result in peaks in the projection spectrum.  Prominent peaks are labeled.  Note that the 1460.82 keV peak (from the $^{40}$K decay) occurs at 1126.85 keV in the 333.97 keV coincidence data, since it is a result of the full energy being shared between both detectors.  }
\label{fig:fullspec}
\end{figure*}

To investigate the coincidence data, events in detector 2 are plotted versus the events in detector 1.  Only the events which meet the coincidence timing requirement (about 4 $\mu$s) between the two detectors and the anti-coincidence timing requirement (about 10 $\mu$s) between the detectors and the veto are plotted in this spectrum (see Fig. \ref{fig:2Dspectrum}).  A few features in this graph are the diagonal lines and the horizontal and vertical lines.  The diagonal lines are the result of a photon from a strong background peak which deposits part of its energy in one detector, and the remainder in the other detector.  When a slice of the 2D histogram is projected onto its parallel axis, this diagonal line can manifest itself as a peak.  These peaks will be referred to as ``shared-energy peaks''.  The horizontal and vertical lines are the result of true coincidences between a strong background peak and a Compton scattered gamma ray from a member of the associated cascade.  See Fig. \ref{fig:fullspec} for a demonstration of the shared energy peaks' effects in the coincidence spectrum.  

The high-intensity spot in the upper right of Fig. \ref{fig:2Dspectrum} is a pulser which was used to determine the dead time.  The use of the pulser was discontinued after a short time due to the low count rate; the dead time was always less than 0.5\%.  A quick calculation can also verify the rate of accidental coincidences.  The event rate seen by each detector is about 0.25 events/s, and the coincidence timing window is 4 $\mu$s.  The accidental rate can be calculated by multiplying the rate in detector 1, the rate in detector 2, and the timing window width.  This calculation results in a coincidence rate of about 14 events in the entire counting time.  These should be distributed evenly over the entire two-dimensional spectrum, and so, do not contribute to events within the region of interest.  In fact, since some of the events counted in the event rate are true coincidental events, this is an overestimate of the coincidence event rate.

\begin{figure}
\centering
\includegraphics
[width=3.5in]{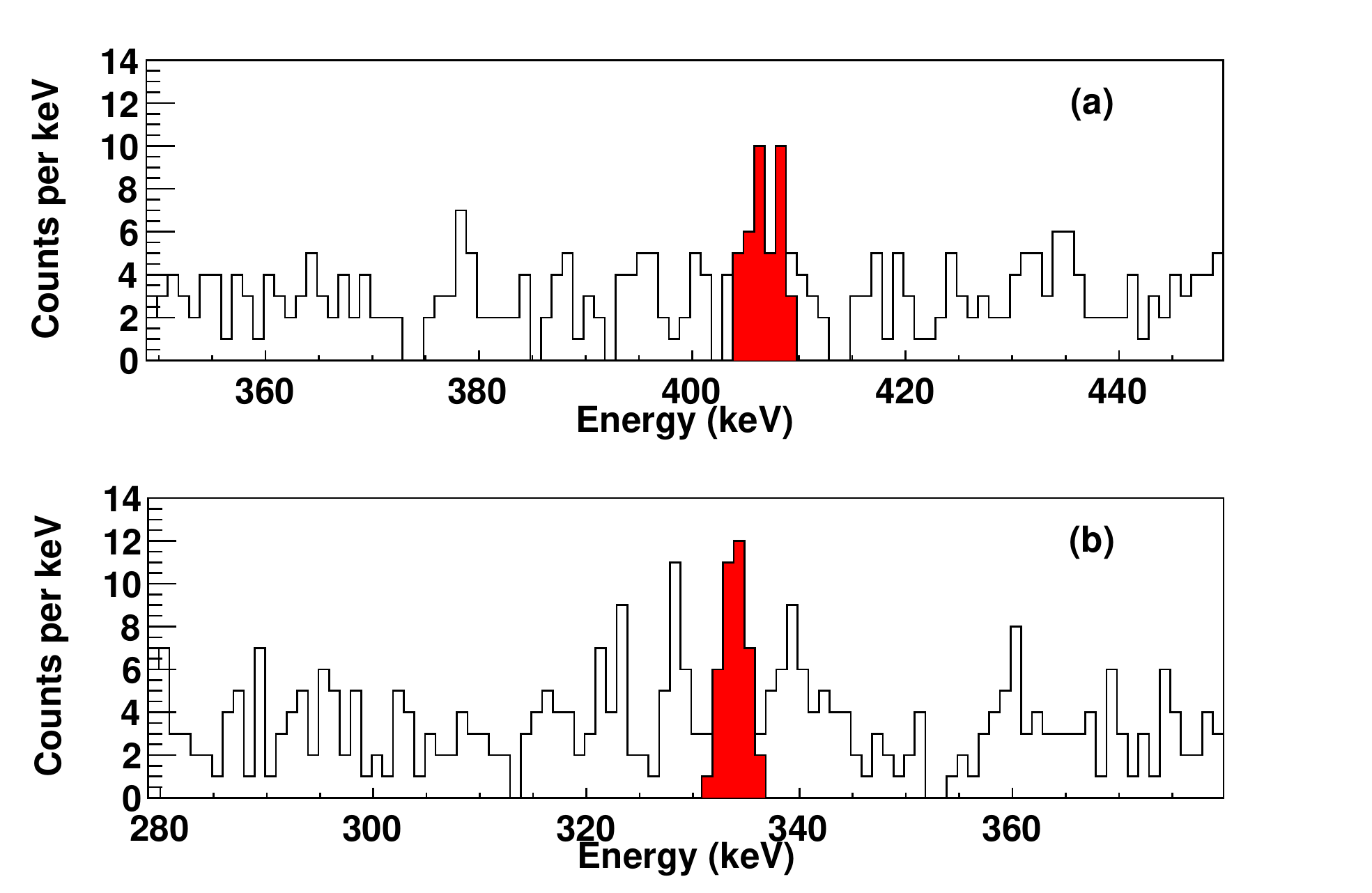}
\caption[Coincidence data for decays to the 0$^+_1$]{(Color online) Coincidence data for decays to the 0$^+_1$ state.  Spectrum (a) is in coincidence with 333.97 keV, and spectrum (b) is in coincidence with 406.52 keV.  The coincidence-energy window in spectra (a) and (b) is $\pm$2.50 keV for detector 2 and $\pm$ 3.00 keV for detector 1.  These values are based on measurements with calibration sources and the background lines referred to in Sec. \ref{sec:analysis}, which also provide information on the pulse-height stability.}
\label{fig:events}
\end{figure}

In the two-dimensional spectrum, the events of interest would occur at coordinates of (333.97 keV, 406.52 keV) and (406.52 keV, 333.97 keV).  The significance of the counts in the two-dimensional regions of interest can best be seen by projecting them onto the x- or y-axis.  To accomplish this, an energy condition ({\it i.e.} 333.97 $\pm$ 2.50 keV) is applied to events in one detector, and all the events which occur in the other detector in coincidence are projected into a histogram.  This results in four histograms; two are in coincidence with 333.97 keV in detectors 1 and 2, and two are in coincidence with 406.52 keV in detectors 1 and 2.  The corresponding histograms can then be summed so that all events in coincidence with a 333.97 keV photon are in one histogram, and vice versa, reducing the number of histograms to two (see Fig. \ref{fig:events}).  To avoid any histogram summing effects, the detectors were carefully gain matched and the bin widths are extremely close.  

\subsection{Coincidence efficiency}\label{sec:coineff}

\begin{figure}
\centering
\includegraphics
[width=4.0in]{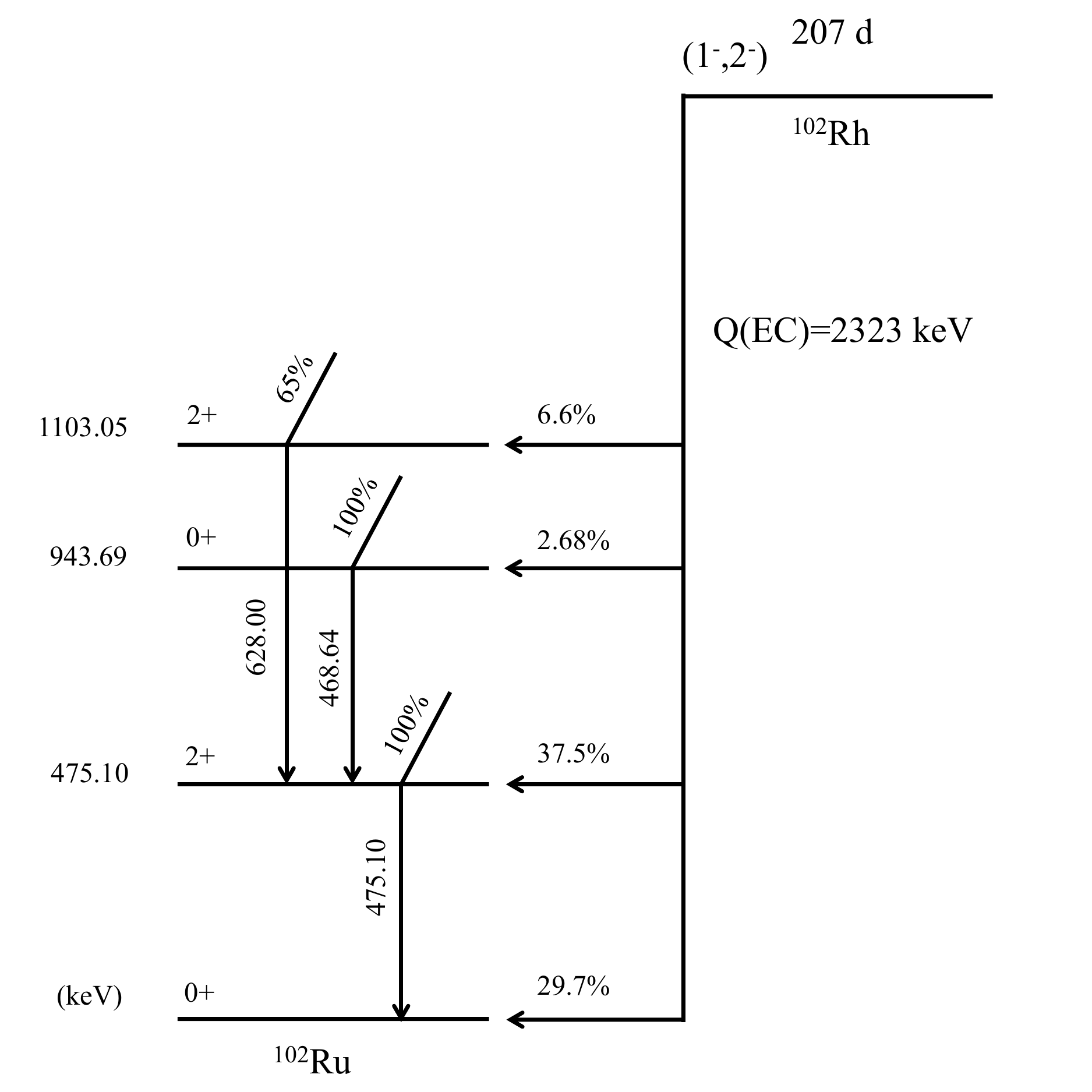}
\caption[Level diagram for the decay of $^{102}$Rh]{The level diagram for $^{102}$Rh is shown.  Note the 0$_1^+\rightarrow 2_1^+\rightarrow 0_{gs}^+$ decay which emits a 468.64 keV gamma ray in coincidence with a 475.10 keV gamma ray.  These were used for coincidence efficiency measurements.}
\label{fig:rh102decay}
\end{figure}

The coincidence efficiency of our double-beta decay apparatus has been measured using a $^{102}$Rh source sandwiched between molybdenum metal disks.  These disks were used for the case of the $^{100}$Mo double-beta decay measurement.  See \cite{Kid09} for details regarding the coincidence efficiency measurement.  Note that this particular source is chosen such that the angular correlation of the decay,
\begin{equation}
\label{equ:angcorr}
W(\theta)=\frac{5}{8}(1-3cos^2(\theta)+4cos^4(\theta)),
\end{equation}
 matches that of the decay of the 0$^+_1$ state of $^{150}$Sm (0$^+\rightarrow$2$^+\rightarrow$0$^+$) and the energies of this decay (468.64 keV -- 475.10 keV) are relatively close to those investigated here (see Fig \ref{fig:rh102decay}).  The data obtained using this method can be scaled for use in the $^{150}$Nd case.  There are two steps to be taken to correct the $^{102}$Rh data to reflect the coincidence efficiency for the present $^{150}$Nd source.  First, by obtaining a fit to the relative efficiency data, the efficiency of either detector at any energy can be determined.  This information is then utilized to scale the 468.64 keV-475.10 keV coincidence yields from $^{102}$Rh to the 333.97 keV-406.52 keV region of interest (ROI).  The yields must be multiplied by this factor:

\begin{equation}
 \frac{\epsilon_{1\gamma}(333.97)\epsilon_{2\gamma}(406.52)+\epsilon_{1\gamma}(406.52)\epsilon_{2\gamma}(333.97)}{\epsilon_{1\gamma}(468.64)\epsilon_{2\gamma}(475.10)+\epsilon_{1\gamma}(475.10)\epsilon_{2\gamma}(468.64)}
 \label{fact1},
 \end{equation}

where the subscripts 1$\gamma$ and 2$\gamma$ refer to the relative efficiency of detectors 1 or 2 at that particular energy.  This factor was calculated to be 1.50$\pm$0.07.

The different attenuation properties of the materials must then be taken into account.  As the original coincidence efficiency measurement was taken for the $^{100}$Mo measurement, molybdenum metal was used as the attenuator.  The attenuator in the present case is Nd$_2$O$_3$.   The attenuation coefficients for 333.97 keV and 406.52 keV were obtained for this material from XCOM, a photon cross-section database online at NIST \cite{NIST}.  This factor is then calculated:

\begin{equation}
 \frac{A_{Nd_2O_3}(333.97)A_{Nd_2O_3}(406.52)}{A_{Mo}(468.64)A_{Mo}(475.10)}=1.839\pm0.010
 \label{fact2},
 \end{equation}
where A$_{Nd_2O_3}$(E$_\gamma$) and A$_{Mo}$(E$_\gamma$) are the survival probabilities of a photon of energy E$_\gamma$ in Nd$_2$O$_3$ and molybdenum metal calculated from the attenuation coefficients.  Note that the attenuation must be calculated with 0.500 $\pm0.005$ cm attenuation length for the molybdenum metal, 0.390$\pm0.005$ cm attenuation length for the Nd$_2$O$_3$.

\begin{table}
\begin{center}
\caption{Summary of systematic error contributions.}
\label{tab:error}
\begin{tabular}{lc} 
\hline
\hline
Uncertainty Contribution  & \% \\
\hline
Intensity of Calibration Gamma Source & 3\%\\
Energy and Attenuation Correction Factors &4.7\%\\
{\it z}-Dependence Correction Factor & 1\%\\
Geometry of $^{150}$Nd Source & 3\%\\
Non-Symmetrical Efficiency Curve & 2.4\%\\
Dead Time & 0.15\%\\
Uncertainty in $^{102}$Rh Half-Life & 0.15\%\\
Total & 6.8\%\\
\hline
\hline
\end{tabular}
\end{center}
\end{table}

\begin{figure}
\centering
\includegraphics
[width=3.3in]{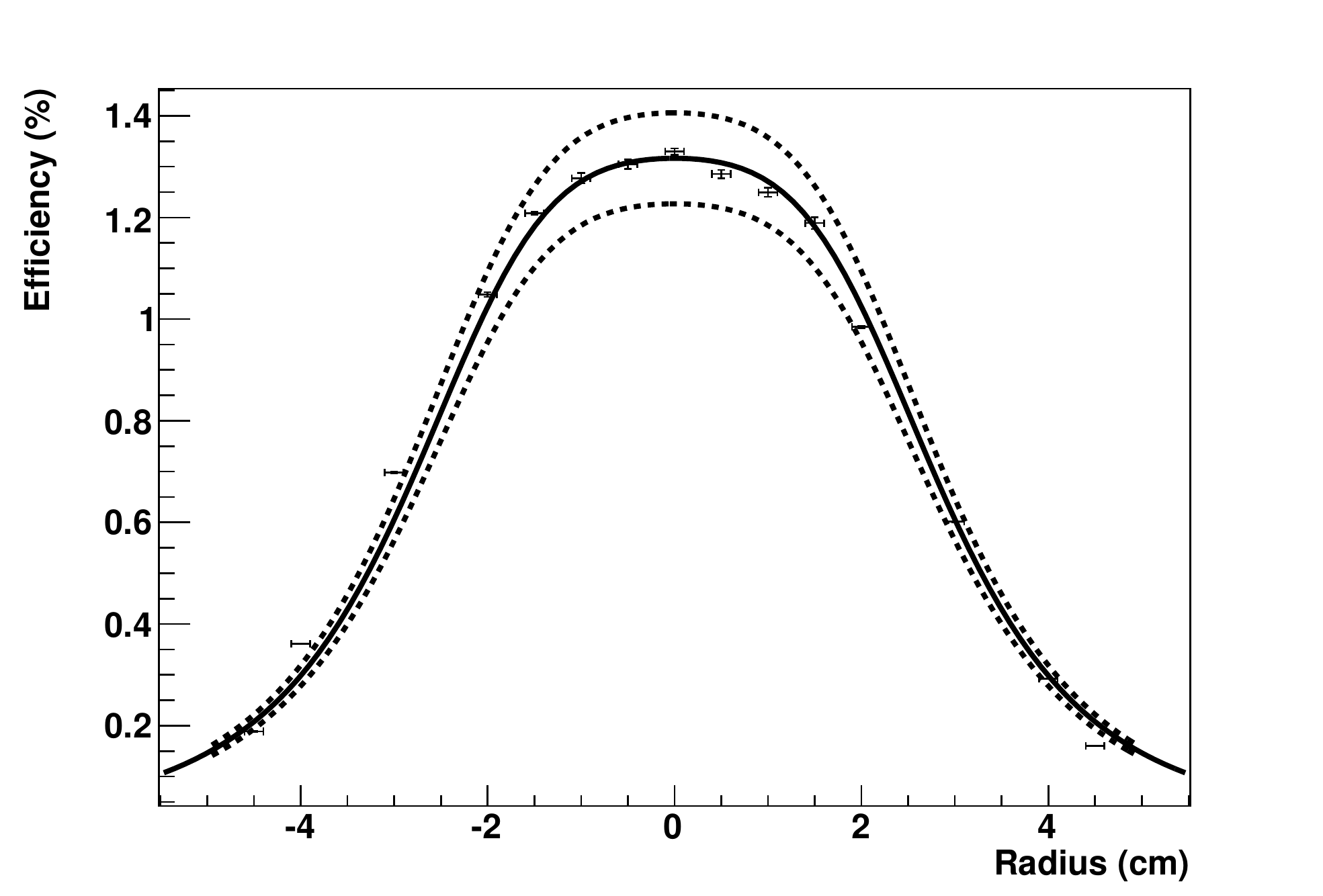}
\caption[Radially dependent coincidence efficiency.]{Coincidence detection efficiency data obtained as a function of $r$ for the E$_{\gamma1}$=333.97 keV and E$_{\gamma2}$=406.52 keV coincidence. The data were taken with the$^{102}$Rh source (E$_{\gamma1}$=468.64 keV and E$_{\gamma2}$=475.10 keV) and then corrected for detection efficiency and attenuation differences between the two $\gamma$--ray pairs involved, and a geometric correction was applied.  The curve through the data presents a least--squares fit. The upper and lower curves indicate the $\pm$6.8\% scale uncertainty associated with our data.}
\label{fig:cavg}
\end{figure}

Figure \ref{fig:cavg} shows the corrected yields for the $^{150}$Nd experiment as a function of radius.  It gives the efficiencies as a function of {\it r}. The error bars represent the statistical uncertainty of the data. The data points are an average of a two scans:  one was horizontally across the face of the detector, and the other was vertically across the face of the detector.  Here, the effect of the position uncertainty of the $^{102}$Rh source is added in quadrature to the statistical uncertainty. The coincidence efficiency is rather small, close to 1.3\% in the -1 cm $<$ {\it r} $<$ 1 cm range and then dropping smoothly to about 0.2\% at {\it r}=4 cm. The size of the cavity of the Nd$_2$O$_3$ holder takes advantage of this effect, concentrating the powder in the most efficient region of the detectors.

The curve through the data points in Fig. \ref{fig:cavg} is a least-square fit using the empirical functional form
 \begin{equation}
 \epsilon_{\gamma\gamma}(r)=\frac{a}{1+br^2+cr^4}  
 \label{eqno2},
 \end{equation}
where $a$, $b$, and $c$ are free parameters and {\it r}=0 refers to the center on the front face of the HPGe detectors. A careful inspection reveals a slight asymmetry in the coincidence efficiency, providing slightly larger values for {\it r}$<$ -2 cm. The lower (dashed) and upper (dashed-dotted) curves shown in Fig. \ref{fig:cavg} represent our systematic uncertainty of 6.8\%.

The radial contribution to the coincidence efficiency was calculated using

 \begin{equation}
 \epsilon_{r}=\frac{2\pi\int\epsilon_{\gamma\gamma}(r)rdr}{2\pi\int rdr},
 \label{eqno3}
 \end{equation}
where $\epsilon_{\gamma\gamma}(r)$ is the best fit obtained from an asymmetric fit to the data. This value is then corrected by the {\it z}-dependence of the coincidence efficiency measured and confirmed by Monte-Carlo simulation \cite{Bar-comm}, where {\it z} is the distance from the face of the detector.  There is a 10\% decrease in efficiency at the center of the 1 cm thick molybdenum disk compared to the front and back faces.  The final results yields a total coincidence efficiency for the $^{150}$Nd geometry of $\epsilon_{tot}=$(1.18$\pm$0.080)\% where the uncertainty of 6.8\% is due to the contributions listed in Table \ref{tab:error}.  They include an estimated 3\% uncertainty due to the slightly irregular shape of our $^{150}$Nd container.

\section{Rejection of Background Candidates}

To correctly interpret the results of the data analysis, a thorough understanding of the types of potential background candidates must be obtained.  In this section, the types of backgrounds which can contribute to the ROI will be discussed.  Specifically, it must be verified that there are no other sources which can produce a 333.97 keV gamma ray in coincidence with a 406.52 keV gamma ray.  

\subsection{Natural decay chains}

\label{sec:naturalbg}
All materials, unless processed to remove them, naturally contain some amount of Potassium (K), Thorium (Th), and Uranium (U).  Radon can also be a worrisome contaminant as it is gaseous and can plate out on surfaces in the detector setup and is a by-product of the natural decay chains.  In the $^{232}$Th series, the radon isotope is $^{220}$Rn, with a half life of 55.6 s.  In $^{235}$U, the radon isotope is $^{219}$Rn with a half life of 3.92 s, and in $^{238}$U, the radon isotope is $^{222}$Rn with a longer half life of 3.82 days.  Because of the $^{232}$Th contamination in the $^{150}$Nd sample, which is discussed later in this section, $^{220}$Rn is definitely present; any naturally occurring contamination from $^{220}$Rn is in addition to the sample contamination.  The natural abundance of $^{235}$U is only 0.72\% compared to the natural abundance of $^{238}$U, which is 99.27\%; therefore, it is unlikely that there will be noticeable contamination from $^{219}$Rn.  

Starting with $^{238}$U, no gamma rays of note are produced until the metastable state of $^{234m}$Pa $\beta^-$ decays to $^{234}$U.  Though the gamma-ray intensities for this decay are universally less than 1\%, there is one gamma-ray emission of interest which occurs with energy 742.81 keV and intensity 0.11\%.  This gamma-ray transition is notable because it can Compton scatter directly into the ROI.  For the $\pm$3.00 keV gate around the ROI, the 742.81 keV gamma ray could scatter with energies between 330.97 and 336.97 keV in coincidence with 411.84 and 405.84 keV, or between 403.52 and 409.52 keV in coincidence with 339.29 and 333.29 keV.  This potential problem will be addressed in Section \ref{sec:ID}.  

From $^{234}$U to $^{214}$Po, many intense gamma rays are emitted, though none within 10 keV of our ROI:  609.32 keV (45.49\%), 768.36 keV (4.895\%), 1120.29 keV (14.92\%), 1238.12 keV (5.83\%), 1764.49 keV (15.30\%), and 2204.06 keV (4.92\%).  However, some very low-intensity gamma-ray energies in the last stage of this decay ($^{214}$Bi$\rightarrow^{214}$Po) are within our regions of interest.  These are:  333.37 keV (0.065\%), 334.78 keV (0.018\%), and 405.72 keV (0.169\%).  Because neither of the transitions accompanied by the gamma rays near 333.97 keV occur in coincidence with the one at 405.72 keV, they are not expected to contribute to the regions of interest.  The remainder of the decays down to the stable $^{206}$Pb do not produce any gamma rays of note \cite{NNDC}.

The largest source of background from a natural decay chain is the $^{150}$Nd sample itself.  This sample was leased from Oak Ridge National Laboratory (ORNL) where it was enriched in the calutrons.  The enrichment process also resulted in a contamination of $^{232}$Th decay products, as confirmed by an independent radioassay.  This contamination manifests itself in a multitude of gamma-ray emissions from the $\beta^-$-decay of $^{228}$Ac to $^{228}$Th.  Three of these characteristic gamma-ray emissions are in close proximity to the ROI of the 333.97-406.52 keV coincidence.  They are 328.00 keV, 338.32 keV, and 409.46 keV with intensities of 2.95\%, 11.27\%, and 1.92\%, respectively \cite{NNDC}.  Again, none of these gamma-ray energies occur in coincidence with each other.  However, the 328.00 keV and the 338.32 keV gammas do appear in the summed event histogram shown in Fig. \ref{fig:events} (b) as excesses of counts to the left and right of the ROI at 333.97 keV.  This feature will be discussed later in Section \ref{sec:ID}.

Further down the $^{232}$Th decay chain, there are other intense gamma rays which should be noted.  The $^{212}$Bi $\beta^-$ decay to $^{212}$Po produces a 727.33 keV gamma ray at 6.67\% intensity which can create a shared-energy peak (see Section \ref{sec:analysis} for a definition of this term) near the ROI.  The isotope $^{212}$Bi can also alpha decay to $^{208}$Tl, whose decay via $\beta^-$ decay to stable $^{208}$Pb produces some very intense gamma rays at 510.77 keV (22.6\%), 583.19 keV (85.0\%), 860.56 keV (12.5\%), and 2614.51 keV (99.75\%).  

This decay chain can also contribute to the areas surrounding the regions of interest via true coincidences which will form peaks in the one-dimensional summed event spectrum.  In Table \ref{tab:bgcoin} these coincidences are listed.  Because the statistical precision is so low, and the intensities of these coincidences often so small, it is difficult to verify that they explain any excesses of counts in the areas around the ROI.

\begin{table}
\caption[Potential coincidences near regions of interest.]{Potential coincidences which could contribute near the ROI. }
\begin{center}
\begin{tabular}{c|c|c|c|c} 
\hline
\hline
Decay                           & ROI $\gamma$-Ray  & Intensity & Coin. $\gamma$-Ray   & Intensity  \\
                                &  Energy (keV)  & (\%)      & Energy (keV)  &  (\%)\\
\hline
$^{228}$Ac$\rightarrow^{228}$Th & 332.37        &  0.40     & 399.62                  &  0.029\\
$^{228}$Ac$\rightarrow^{228}$Th & 332.37        &  0.40     & 419.42                  &  0.021\\
$^{228}$Ac$\rightarrow^{228}$Th & 338.32        &  11.27    & 372.57                  &  0.0067\\
$^{228}$Ac$\rightarrow^{228}$Th & 338.32        &  11.27    & 377.99                  &  0.025\\
$^{228}$Ac$\rightarrow^{228}$Th & 338.32        &  11.27    & 399.62                  &  0.029\\
$^{228}$Ac$\rightarrow^{228}$Th & 338.32        &  11.27    & 416.30                  &  0.0132\\
$^{228}$Ac$\rightarrow^{228}$Th & 338.32        &  11.27    & 419.62                  &  0.021\\
$^{152}$Eu$\rightarrow^{152}$Gd & 411.12        &  2.237    & 344.28                  &  26.6\\
$^{214}$Bi$\rightarrow^{214}$Po & 405.72         &  0.169    & 304.2                   &  0.019\\
$^{214}$Bi$\rightarrow^{214}$Po & 405.72         &  0.169    & 314.9                   &  ?\\
\hline
\hline
\end{tabular}
\end{center}
\label{tab:bgcoin}
\end{table}

There are two other contaminants to the sample which can be seen in the two-dimensional spectrum, and in one case, contribute to the background near the ROI.  The first of these is $^{176}$Lu, which $\beta^-$ decays to $^{176}$Hf.  Lutetium is a rare-earth metal which forms the same oxide structure, Lu$_2$O$_3$, as Neodymium.  The natural abundance of the isotope $^{176}$Lu is 2.59\% and it has a half life of 3.76$\times$10$^7$ years.  The transition to $^{176}$Hf does not emit gamma rays which contribute in the ROI, but for neutrinoless double-beta decay experiments which may use enriched $^{150}$Nd in the future, the potential contamination by $^{176}$Lu might be of interest. 

The second rare-earth contaminant which has been found in this $^{150}$Nd sample is $^{152}$Eu.  Again, Europium forms the same oxide structure as Neodymium, Eu$_2$O$_3$, and $^{152}$Eu $\beta^-$ decays to $^{152}$Gd and emits (among others) a gamma ray at 344.28 keV (26.6\%) in coincidence with a gamma ray at 411.12 keV (2.24\%).  The limits which define the 406.52 keV ROI cut into the tail of this coincidence which appears in the two-dimensional spectrum.  This seems to account for an excess of counts near 344 keV.  

\subsection{Identification of natural background in data}
\label{sec:ID}

Because the background discussed in the previous section cannot ever be completely removed, its contribution in and around the ROI must be understood.  Mostly, these decays contribute in a general way to the Compton continuum, but a few of the gamma-ray emissions make contributions in the one-dimensional histograms discussed in Section \ref{sec:analysis}.  The first specific natural background lines to be discussed are the 742.81 keV gamma ray emitted in the decay of $^{234m}$Pa, the 768.36 keV gamma ray emitted in the decay of $^{214}$Bi, and the 727.33 keV gamma ray emitted in the decay of $^{212}$Bi.

\begin{figure}
\begin{center}
\includegraphics
[width=3.5in]{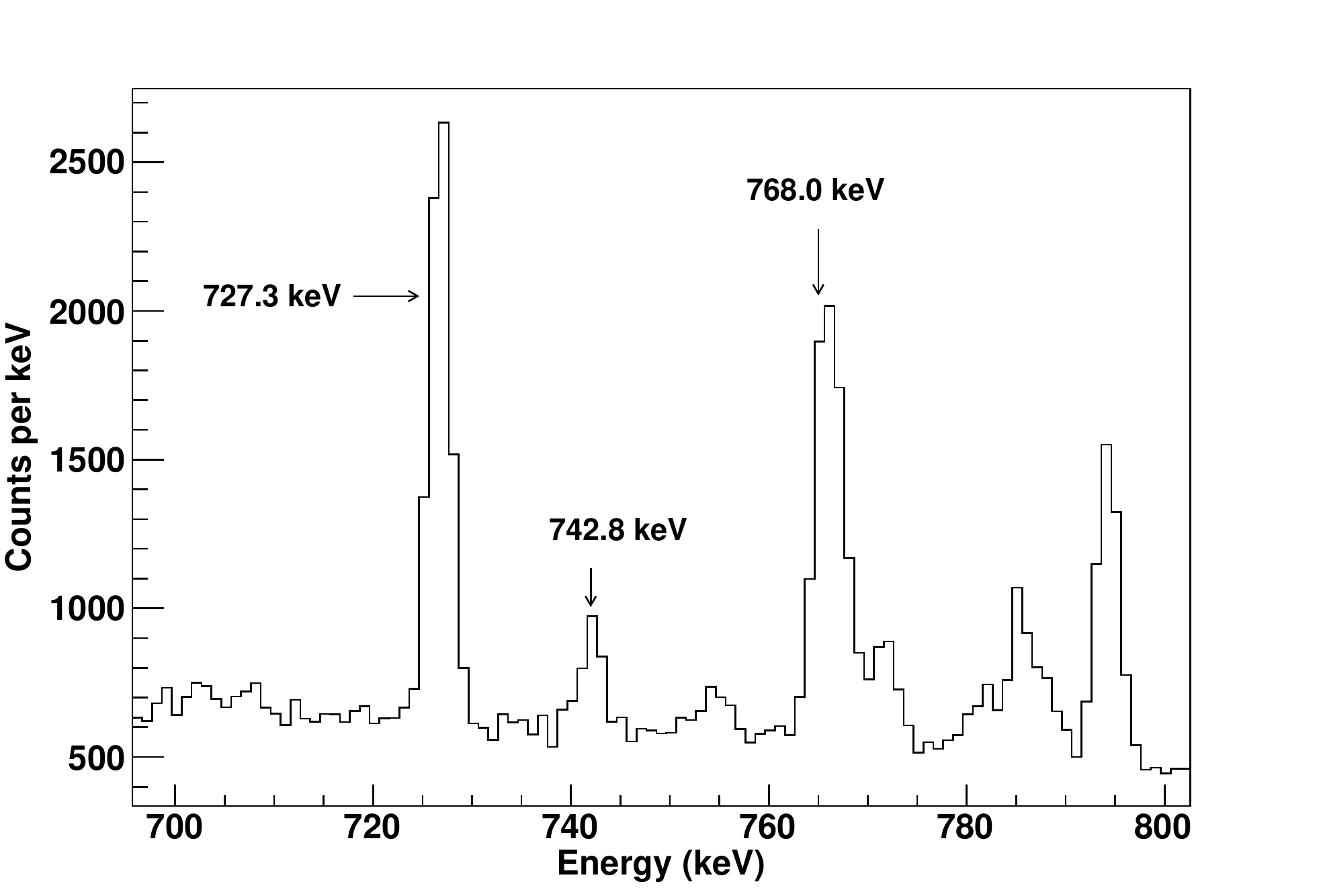}
\caption[Comparison of 742.81 keV to surrounding peaks.]{A comparison of the 742.81 keV gamma-ray peak from the decay of $^{234m}$Pa to surrounding peaks.}
\label{fig:743comp}
\end{center}
\end{figure}

\begin{figure}
\centering
\includegraphics
[width=3.8in]{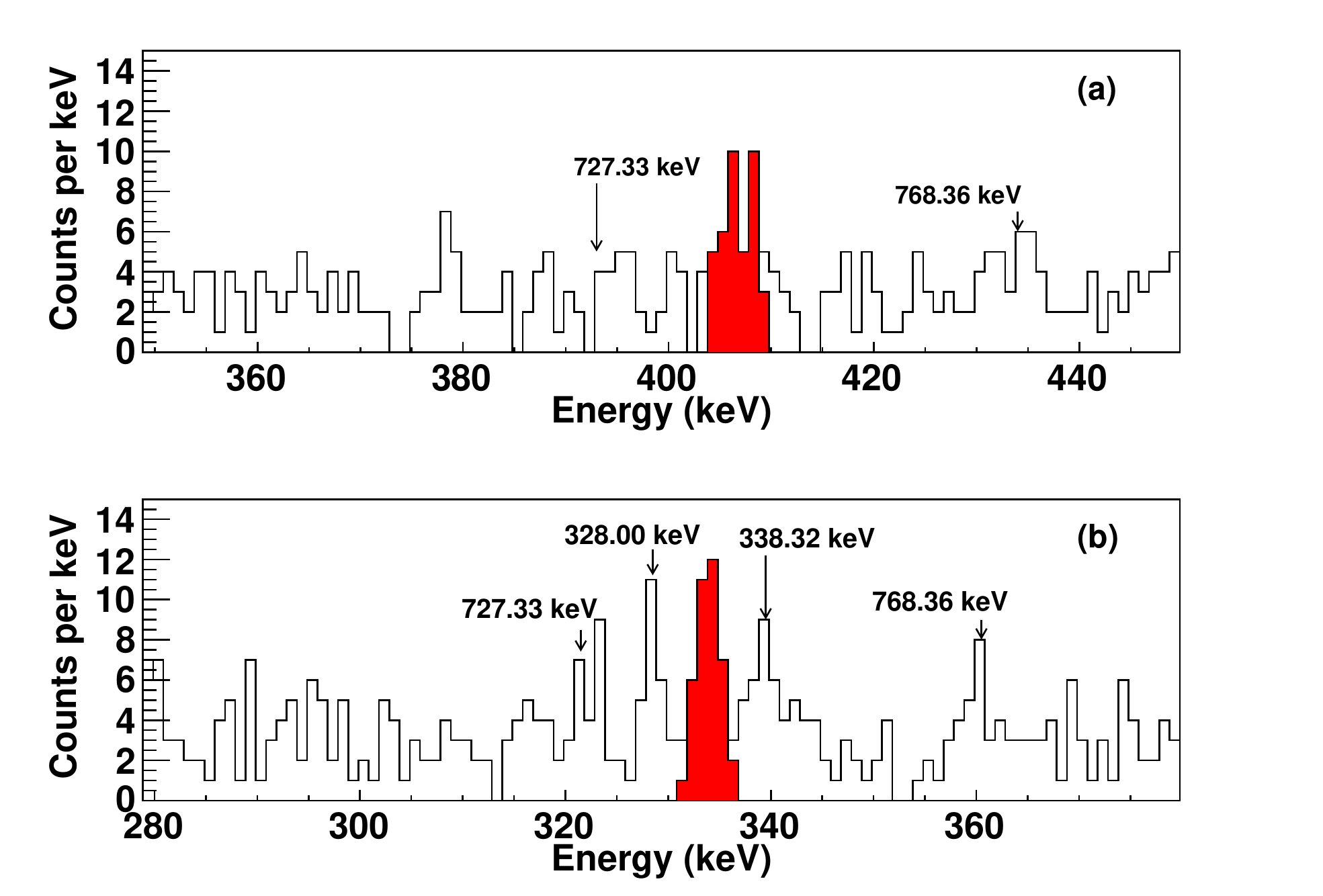}
\caption[The projected and summed event histograms with shared-energy peaks labeled.]{(Color online)  Same as Fig. \ref{fig:events} with nearby background peaks labeled.  The origin of the peaks at 328.00 keV and 338.32 keV in Fig. (b) are discussed in Sec. \ref{sec:naturalbg} and the origin of the shared-energy peaks at 727.33 keV and 768.36 keV in Fig. (a) and (b) are discussed in Sec. \ref{sec:ID}.  }
\label{fig:743coin}
\end{figure}

Examining the singles spectrum near these energies, it can immediately be observed that the 742.81 keV full-energy peak is substantially smaller than the surrounding 727.33 keV and 768.36 keV full-energy peaks (see Fig. \ref{fig:743comp}).  If the one-dimensional coincidence spectrum is then examined, the shared-energy peaks attributed to the 727.33 keV and 768.36 keV $\gamma$ rays can be identified.  By comparing the intensity of the shared-energy peaks to the intensity of the full-energy singles peaks, it can be estimated that the 742.81 keV $\gamma$ ray would  contribute 1.2 $\pm$ 0.6 counts in the ROI over the entire counting period (see Fig. \ref{fig:743coin}).  

The other background which contributes in the areas surrounding the ROI for 333.97 keV are the gamma rays emitted in the decay of $^{228}$Ac to $^{228}$Th which have energies of 328.00 keV and 338.32 keV.  In the two-dimensional spectrum near the regions of interest, these gamma-ray transitions are responsible for creating vertical and horizontal lines at the aforementioned energies which extend up to 1400 keV.  As discussed in Section \ref{sec:analysis}, these are formed when the full energy of a gamma-ray transition is deposited in one detector in coincidence with a Compton-scattered gamma ray in the other detector which does not deposit its full energy.  The effect of the one-dimensional summed event spectrum is to take a cross section of the two-dimensional energy spectrum.  Thus, when the cross section is taken across one of these horizontal or vertical lines, an excess of counts in the one-dimensional spectrum can occur.  This is the cause of the ``peaks" which occur noticeably at 328 keV, and less definitely at 338 keV, as can be seen in Fig. \ref{fig:events} (b).  

Two issues arise from this realization that the cross section of Compton scattering coincidences can create peaks in the one-dimensional summed event spectrum.  The first of these is whether a peak can be expected in the other one-dimensional spectrum from a Compton-scattering coincidence with the 409.46 keV peak in the decay of $^{228}$Ac to $^{228}$Th.  The second of these also helps in the discussion of the previous issue, and is related to the intensities of the Compton-scattering coincidence peaks.  As stated previously, the intensities of the full-energy peaks for 328.00 keV, 338.32 keV, and 409.46 keV are 2.95\%, 11.27\%, and 1.92\%, respectively, but the Compton-scattering coincidence peaks do not appear to follow these intensities.  To understand the contribution of the 409.46 keV Compton scattering, the intensities must be calculated using the branching ratios to all states which make a transition through the 328.00 keV level, the 396.08 keV level (through which transition results in the 338.32 keV gamma ray), and the 1431.98 keV level (through which transition results in the 409.46 keV gamma ray).  It can be calculated from the NNDC data tables \cite{NNDC} that 2.9\% of all $^{228}$Ac decays to $^{228}$Th proceed in coincidence with 328.00 keV with energies large enough to Compton scatter into the 406.52 keV ROI.  Similarly, the percentage of decays in coincidence with 338.32 keV is 2.3\%, and for 409.46 keV, only 0.26\%.  Thus, it makes sense that the excesses of counts in the 328.00 keV and 338.32 keV areas are of a comparable size, and it can be deduced that only one-tenth of those counts are expected at 409.46 keV.  

\begin{figure}
\begin{center}
\includegraphics
[width=3.5in]{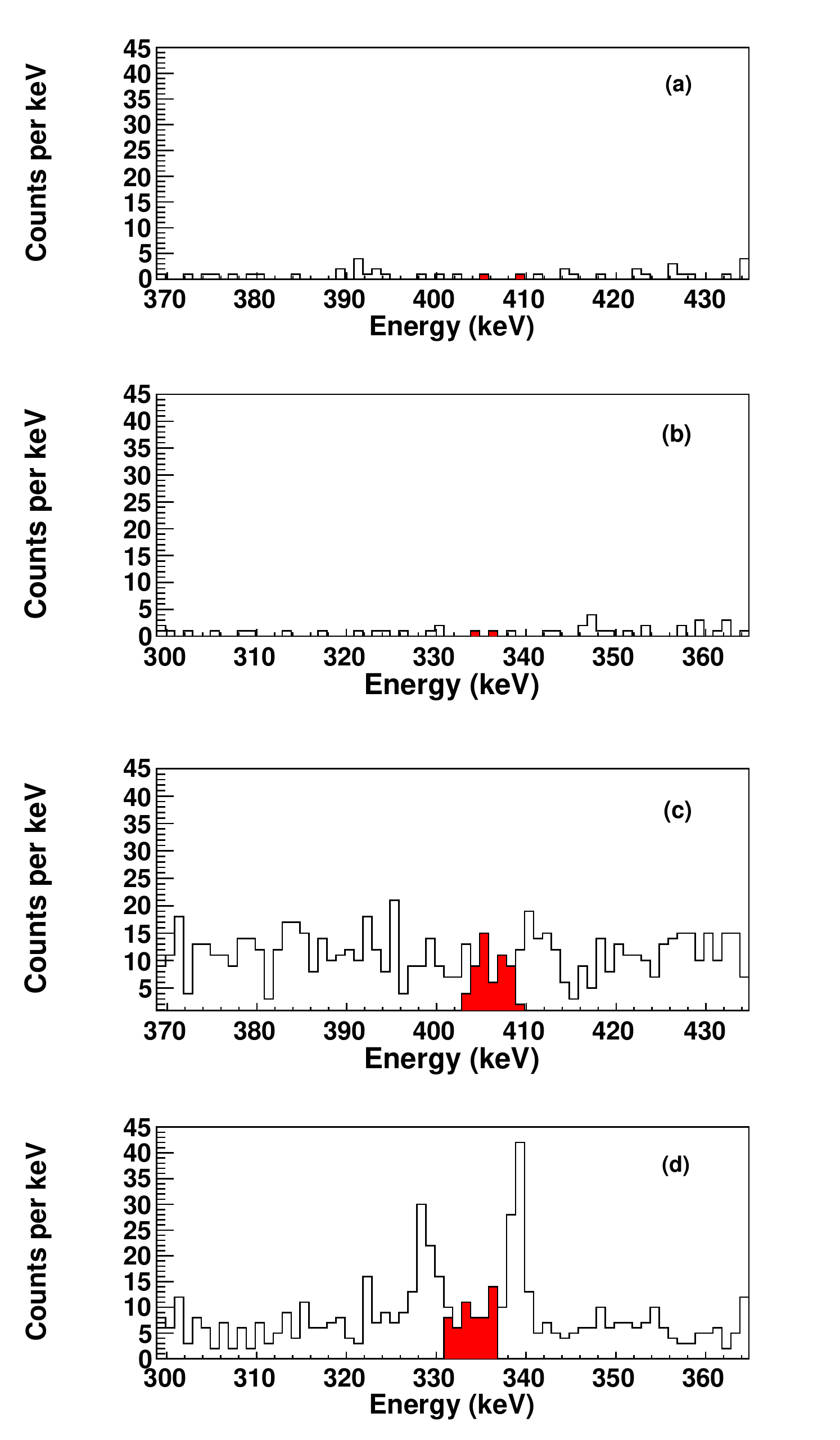}
\caption[Background data and thorium oxide data.]{(Color online) Coincidence data shown for the background runs taken over 180.76 days with no sample inserted are shown in spectra (a) and (b).  Coincidence data shown for the ThO$_2$ data taken over 1.10 days are shown in spectra (c) and (d).  In all spectra, the coincidence window is $\pm$ 3.00 keV.}
\label{fig:bg-th}
\end{center}
\end{figure}

To further explore this source of background, two further data sets were acquired.  One data set is taken over a time period of 180.76 days with no sample inserted between the detectors (see Figs. \ref{fig:bg-th} (a) and (b)).  The drastic effect on the background can immediately be seen.  The time period over which this is taken is equal to about 28\% of the time period over which the $^{150}$Nd was taken.  This confirms that the background in the sample clearly dominates over backgrounds in the experimental environment.  

The second data set was intended to study the gamma-ray lines which surround the ROI and how the thorium contamination contributes in the ROI.  A ThO$_2$ sample was placed between the detectors and counted for 1.10 days (see Figs \ref{fig:bg-th} (c) and (d)).  This data set was analyzed in two ways.  First, the analysis on the $^{150}$Nd data, which will be further described in Section \ref{DBBDgs}, was identically performed on this ThO$_2$ data set.  This analysis revealed zero excess counts in the ROI compared to the continuum.  The second analysis performed on the ThO$_2$ data compared the areas of the 328.00 keV and 338.32 keV peaks to the area inside the ROI.  This ratio was then applied to the $^{150}$Nd data to discover how many events are expected in the ROI due to the thorium contamination.  Since it was already verified that the background is dominated by this contamination, this result should be the sole contribution to events in the ROI.  This analysis gave the same results which will be presented in Section \ref{DBBDgs}.  Therefore, we can be confident that the thorium contamination is not responsible for any excess of counts in the ROI.  

\subsection{Cosmic-ray backgrounds}
\label{sec:cosmicray}

For a low-background counting system such as our double-beta decay setup, cosmic rays could account for a significant portion of the background.  Fast neutron flux at sea level is the result of the muons interacting with high-Z materials.  The neutrons can then interact in materials via (n, p), (n, $\alpha$), (n, 2n), and (n,n'$\gamma$).  Reducing this background contribution can be done by going underground.  By installing our double-beta decay setup at KURF, the muon flux was reduced by a factor of 4$\times10^5$.  

Cosmic-ray background could permeate the regions of interest via proton and neutron reactions on nuclei near or within the source, resulting in production of the excited state  in the final nucleus which would appear like a double-beta decay event.  It must be shown that the events seen in the ROI were not produced by any other mechanism than double-beta decay.  

\subsubsection{Proton activation}

A potential background candidate results from proton activation of the $^{150}$Nd nucleus.  A significant proton flux through the apparatus could result in the (p,n) activation of $^{150}$Nd, producing the nucleus which would be the intermediate state in double-beta decay, $^{150}$Pm.  $^{150}$Pm can then $\beta$ decay to $^{150}$Sm with a half-life of 2.68 hours.  This $\beta$ decay can produce the excited-state decays which would be attributed to a double-beta decay event.  Because the half life of the activated nucleus is so short, any activation due to a significant proton flux, such as during transport would have died away long ago, and so the only concern in this case would be continual activation during the experiment.  

Though the cosmic-ray flux has been significantly reduced by going underground, because of the extremely long half life of double-beta decay, even a small flux is reason for concern, although it is unlikely that a primary cosmic-ray proton can penetrate the overburden at KURF.

\begin{figure}
\begin{center}
\includegraphics
[width=3.5in]{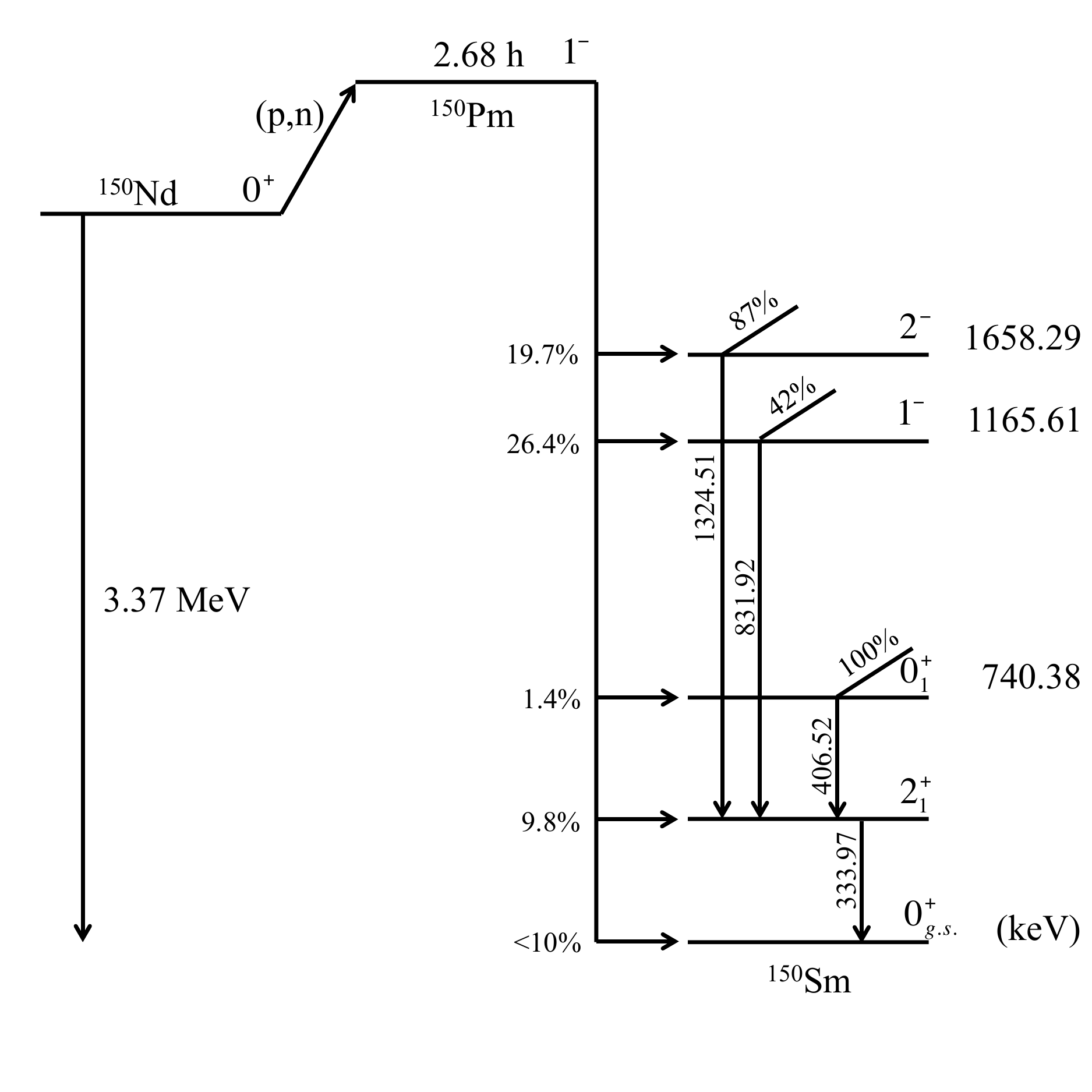}
\caption[Beta-decay scheme of $^{150}$Pm.]{The decay scheme of $^{150}$Pm $\beta$ decay after (p,n) activation of $^{150}$Nd.}
\label{fig:nd150pn}
\end{center}
\end{figure}

To investigate the (p,n) case, the $\beta$ decay of $^{150}$Pm must be studied.  As can be seen in Fig. \ref{fig:nd150pn}, only 1.4\% of the $\beta$ decays feed the 0$^+_1$ excited state.  It should also be noted that 19.7\% of the $\beta$ decays feed the 2$^-$ state at 1658.29 keV, and 26.4\% feed the 1$^-$ state at 1165.61 keV.  Both of these states decay emitting a 1324.51 keV gamma or a 831.92 keV gamma, respectively, in coincidence with the 333.97 keV gamma.  The probability of these decays are much larger than the probability of the decay to the 0$^+_1$ excited state, so searches for these higher excited-state coincidences will reveal the possible contribution to the double-beta decay regions of interest.  Indeed, no coincidences were observed above background at either 1324.51-333.97 keV or 831.92-333.97 keV.  As a result, a limit on the rate of events in the ROI was calculated to be $<2.5\times10^{-9}$ Hz, or a contribution of $<$0.14 counts in the ROI over the entire counting time.  Thus it is extremely unlikely that (p,n) activation of $^{150}$Nd contributes to the background in the regions of interest.  

\subsubsection{Neutron background}

Finally, a significant neutron flux through the detectors and source disk provides another potential source of background.  The lead brick enclosure shown in Fig. \ref{fig:setup} is not very effective in stopping neutrons; the cross section for neutron capture is rather small.  The source of a neutron flux could be due to primary cosmic-ray flux, or from secondary interactions such as (p,n) occurring in the room or in the setup.  Neutrons can be produced by high-energy protons interacting in the lead or via ($\alpha$, n) reactions, especially in the glass of the PMTs used in the NaI annulus.  A variety of interactions are available once a neutron flux is present, such as (n,$\gamma$) capture, (n,n'$\gamma$) inelastic scattering, elastic scattering, or even (n, 2n).  

The presence of a neutron flux can be checked by examining the $\gamma$-ray energy singles spectrum for characteristic neutron interactions with germanium nuclei.  Neutrons which scatter inelastically off germanium nuclei in the crystal generate peaks which are broadened due to the recoil of the nucleus.  Referring again to Fig. \ref{fig:zoom}, when our double-beta decay setup was operated at the LBCF, at 595.85 keV and at 689.6 keV there are peaks with the distinctive sawtooth shape originating from (n,n'$\gamma$) on $^{74}$Ge (which is 36.5\% naturally abundant) and $^{72}$Ge (27.4\%).  Specifically referring to the latter of these two interactions, it was occurring at a rate of 20.3 day$^{-1}$.  However, once underground at KURF, the signal is indistinguishable from background, setting a rate limit of $<$5.4  day$^{-1}$.   Since none of these interactions result in coincident gamma rays in our region of interest, this reduction is not necessarily applicable to our setup, but is purely a matter of interest.  

\section{Double-Beta Decay of $^{150}$N\lowercase{d} to 0$_1^+$} \label{DBBDgs}

As stated in Section \ref{sec:analysis}, obtaining the raw data is a straightforward procedure following the accumulation of coincidence data.  There are no complex data processing or analysis routines.  The strength of the data lies in its simplicity; the data is compiled over the entire counting period, and any positive peaks in or near the ROI must be identified.  

In the search for double-beta decay of $^{150}$Nd to the 0$^+_1$ state (740.38 keV) of $^{150}$Sm, data were accumulated over a total of 642.8 days.  Recall that identification of this decay involves the simultaneous detection of the coincidence de-excitation (0$^+_1\rightarrow$2$^+_1\rightarrow$0$_{gs}$) gamma-ray photons with energies of 333.97 keV and 406.52 keV.  The branching ratio for this cascade is 100\%.  Gamma-ray energy spectra in coincidence with 333.97 keV and 406.52 keV, presented in Fig. \ref{fig:events} show the detection of the coincidence events.  There are two coincidence widths shown in Fig. \ref{fig:events}.  The width of the coincidence window is $\pm$2.50 keV for detector 2 (due to its slightly better resolution and the excess of $^{228}$Ac decays seen by this detector), and $\pm$3.00 keV for detector 1.  Though there are other peaks near the ROI, the excess of counts at the correct energy is undeniable.  The significance of the events in the ROI depends on the identification of the surrounding background peaks.  

The half-life calculation can be accomplished through the use of the definition in Equation \ref{equ:halflife},

\begin{equation}
\label{equ:halflife}
T_{1/2}=\frac{ln2N_0t\epsilon_{\gamma\gamma}^{tot}}{N_{\gamma\gamma}}.
\end{equation}
Here, $N_0$ is the number of $^{150}$Nd nuclei in the sample, $t$ is total counting time (642.8 days or 1.76 years), $\epsilon_{\gamma\gamma}^{tot}$ is the total coincidence detection efficiency for the 333.97-406.52 keV cascade, and $N_{\gamma\gamma}$ is the number of detected coincidence events minus background.  The value of $N_0$ is 1.612$\times$10$^{23}$ nuclei of $^{150}$Nd (see Sec. \ref{sec:method}).  As mentioned in Section \ref{sec:coineff}, the coincidence efficiency for the 333.97-406.52 keV cascade is $\epsilon_{\gamma\gamma}=$(1.18$\pm$0.080)\%.  

Strictly speaking, in evaluation of background, existing peaks should be removed from the spectrum before background integration.  To this end, effort must be made to identify all background peaks surrounding the regions of interest.  In low-background counting experiments, this is a difficult endeavor due to the small number of counts involved.  The evaluation of background is concentrated in regions where the continuum is perceived to be flat.  In Fig. \ref{fig:events} (a) it is apparent that there is a peak located at 377.99 keV which can be attributed to a coincidence between gamma rays emitted from a cascade from the 1531.47 keV level in $^{228}$Th.  Thus, it is excluded from the background integration.  In Fig. \ref{fig:events} (b) the region from 315 keV to 345 keV was excluded from the background integration.  These background integrations result in an average background of 3.0 counts per keV.  In spectra (a) and (b) in Fig. \ref{fig:events} the average ROI is taken to be 5.5 keV wide.  There are 39 raw events in (a) and (b).  Aside from the regions mentioned, the background was integrated from 350.00 keV to 445.00 keV in Fig. \ref{fig:events} (a) and 280.00 keV to 350.00 keV in Fig. \ref{fig:events} (b).  After subtracting the background from the raw events and the single count discussed in Sec. \ref{sec:ID}, an average of 21.6 $\pm$ 6.4 events was obtained.  Inserting this into Equation \ref{equ:halflife}, the following value is obtained.
 
\begin{equation}
\label{equ:finalanswer}
T_{1/2}=(1.07^{+0.45}_{-0.25}(stat)\pm0.07(syst.))\times 10^{20} \text{ years}.
\end{equation}

From the value for T$_{1/2}$, the nuclear matrix element for this particular transition can also be calculated.  Using the value of G$^{2\nu}=4329\times10^{-21}$ years$^{-1}$ for the 0$^+_1$ state from \cite{Kot12}, the effective NME, $|M_{eff}^{2\nu}|$, can be extracted.  $|M_{eff}^{2\nu}|$ contains factors of the axial vector coupling constant, $g_A$, and the rest mass of the electron, $m_ec^2$.  Here, $|M_{eff}^{2\nu}|$ is found to be 0.0465$^{+0.0098}_{-0.0054}$.  

Comparing this value obtained for $|M_{eff}^{2\nu}|$ for the 2$\nu\beta\beta$ decay of $^{150}$Nd to the first excited 0$^+_1$ state of $^{150}$Sm to the one from \cite{Kot12} to the 0$^+_{gs}$ ground state of $^{150}$Sm, we find that the ratio of the nuclear matrix elements is M$^{0\nu}(0^+_1)/M^{0\nu}(0^+_{gs}$)=0.80$\pm$0.14. This observation is at variance with the calculations of \cite{Bel13} for the 0$\nu\beta\beta$ decay of $^{150}$Nd, which predict a ratio of 1.6.


\section{Double-Beta Decay of $^{150}$N\lowercase{d} to Higher Excited States}

\begin{figure*}
\centering
\includegraphics
[width=7in]{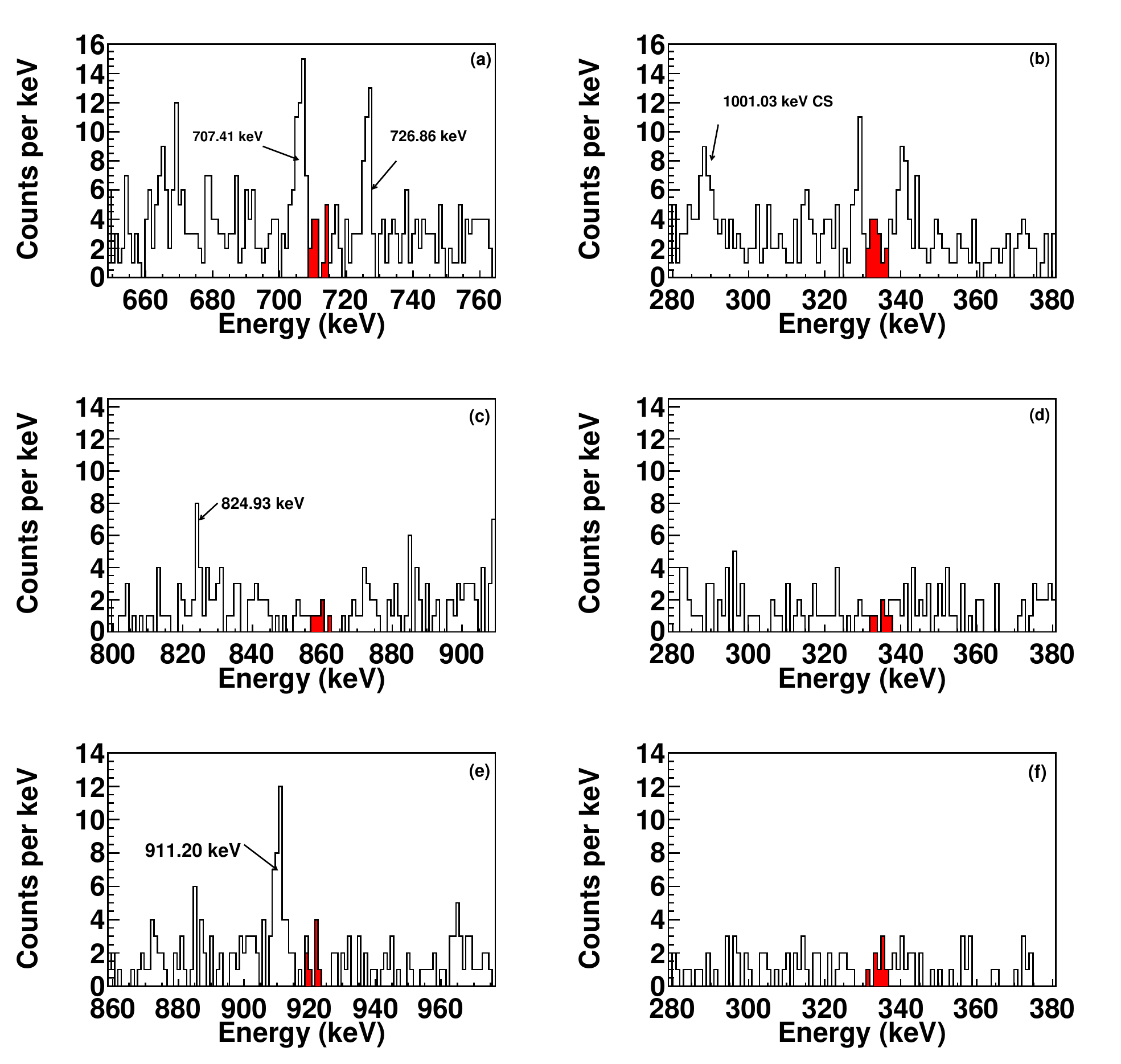}
\caption[Coincidence data for higher excited states.]{(Color online) Coincidence data for the higher excited states.  Spectra (a) and (b) are in coincidence with 333.97 keV and 712.21 keV, respectively, (c) and (d) are in coincidence with 333.97 keV and 859.87 keV, respectively, and (e) and (f) are in coincidence with 333.97 keV and 921.2 keV, respectively.  The coincidence-energy window is $\pm$ 3.00 keV.}
\label{fig:events-higher}
\end{figure*}

The double-beta decay of $^{150}$Nd to higher excited states of $^{150}$Sm was also studied.  Following Fig. \ref{fig:nd-higher}, the coincidence energies for these three cascades are 712.21 keV - 333.97 keV, 859.87 keV - 333.97 keV, and 921.2 keV - 333.97 keV.  Note that the 712.21 keV transition and the 859.87 keV transition are both 2$^+\rightarrow2^+\rightarrow0^+$ decays, and the efficiency must be adjusted accordingly.  As seen in Fig. \ref{fig:events-higher}, no events were detected above background for any of these decays, meaning that half-life limits may be set for double-beta decay to these states.  The half-life limit is given by

\begin{equation}
\label{equ:halflifelimit}
T_{1/2}>\frac{ln2 N_0 t f_b \epsilon_{\gamma\gamma}^{tot}}{N_d},
\end{equation}
where $f_b$ is the branching ratio for the particular cascade.  $N_d$ is a value chosen as recommended by the Particle Data Group \cite{PDG}.  It is determined by a statistical estimator which is derived for a process obeying Poisson statistics.  The definition of $N_d$ is given as the desired upper limit on the signal above background, and it depends on the chosen level of certainty.  For double-beta decay half-life limits, reporting the 90\% confidence level (CL) has become standard.  For the 90\% CL, $N_d$ is the maximum signal which will be observed in 90\% of all random repeats of the experiment, or there is a less than 10\% chance of observing a count rate above $N_d$.  The results for the decays to the higher excited states of $^{150}$Sm are summarized in Table \ref{tab:summary}.

\section{Conclusions}

The results from the counting of $^{150}$Nd for 642.8 days by our double-beta decay setup at Kimballton Underground Research Facility have been presented.  These results are the culmination not only of two years of experimental counting time, but also several years of preparation and establishment of TUNL's first remote underground laboratory.  The significance of this result will now be discussed.  

The major result from this experiment is the observation of the 333.97 keV - 406.52 keV coincidence which confirms the double-beta decay of $^{150}$Nd to the 0$^+_1$ state of $^{150}$Sm.  Though this decay has been observed recently by Barabash {\it et al.} \cite{Bar09}, this is the first measurement which confirms that the coincidence has taken place.  The results in \cite{Bar09} were obtained using a single high-purity germanium (HPGe) detector whose endcap was surrounded by 3046 g of natural Nd$_2$O$_3$.  The total exposure was 1732 kg-h.  These data do not take advantage of the coincidence technique used in the present work and so have significantly higher background in the regions of interest.  Furthermore, Barabash {\it et al.} must take into account the numerous $\gamma$-ray lines which contaminate the regions of interest which are listed in Table \ref{table:nd150bg}.  None of those have coincident partners in the regions of interest and are therefore not likely to enter our spectra.  

\begin{table}[t]
\caption[Background contamination in $^{150}$Nd regions of interest]{The most likely candidates for contamination in the 333.97 keV and 406.52 keV regions of interest for $^{150}$Nd.}
\begin{center}
\begin{tabular}{c|c|c|c} 
\hline
\hline
Isotope & Intensity  & Energy  & Decay\\
        & (\%)       & (keV)   & Chain\\
\hline
$^{214}$Bi & 0.065 & 333.37 & $^{238}$U\\
$^{214}$Bi & 0.036 & 334.78 & $^{238}$U\\
$^{214}$Bi & 0.169 & 405.72 & $^{238}$U\\
$^{228}$Ac & 0.40 & 332.37  & $^{232}$Th\\
$^{227}$Th & 1.54 & 334.37  & $^{238}$U\\
$^{211}$Pb & 3.78 & 404.85 & $^{238}$U\\
\hline
\hline
\end{tabular}
\end{center}
\label{table:nd150bg}
\end{table}

Though the setup of Barabash {\it et al.} is rather simple, employing only one HPGe detector, there are still inherent difficulties to making a low-background measurement.  Because of the coincidence mechanism used in the present work, there were many potential background candidates which could be discarded from the start.  These background candidates continue to plague the measurement made by Barabash {\it et al.} \cite{Bar09}, who rely on observation of the 333.97 keV or 406.52 keV peak in a singles spectrum.  The reliability of this result is dependent upon the fit to the background spectrum and the simulation of the potential contaminants in the regions of interest.  Their result, $T_{1/2}=1.33^{+0.36}_{-0.23}(stat.)^{+0.27}_{-0.13}(syst.)\times 10^{20}$ years does agree within error with the present result, $T_{1/2}=(1.07^{+0.45}_{-0.25}(stat)\pm0.07(syst.))\times 10^{20}$ years.  It is interesting to point out that our result obtained by using the coincidence technique has about the same statistical uncertainty as the result of \cite{Bar09} based on a singles measurement with almost a factor of three larger exposure (1732 kg-h versus 619 kg-h in the present case).  In addition, our systematic uncertainty is considerably smaller than that of Ref. \cite{Bar09}.  The recent systematic analysis of Ren and Ren \cite{Ren14} gives the value T$_{1/2}$=0.776$\times$10$^{20}$ years, in closer agreement with our result than that of Ref. \cite{Bar09}.The observation of double-beta decay to an excited final state has only been previously observed in one other nucleus, $^{100}$Mo.  (\cite{Kid09} and references therein) 

The limits for the decay of $^{150}$Nd to the excited states of $^{150}$Sm are summarized in Table \ref{tab:summary} and include limits determined by Barabash {\it et al.} \cite{Bar09}.  The limits determined in \cite{Bar09} range from about an order of magnitude to a factor of 3 better than the limits found in the present work.  In these cases, our small coincidence efficiency makes singles measurements more favorable.    

In Ref. \cite{Suh98}, \cite{Aun96}, and \cite{Gri92} a different dependence of the particle-particle strength parameter ($g_{pp}$) was found in calculations of T$_{1/2}$ for 2$\nu\beta\beta$ decay transitions to the ground state and excited final states.  It would be very interesting to see whether the calculations of Fang {\it et al.} for $^{150}$Nd support this observation.  If confirmed, the present datum of T$_{1/2}$ for the 2$\nu\beta\beta$ decay of $^{150}$Nd to the 0$_1^+$ state in $^{150}$Sm could be used not only to fine-tune theoretical calculations, but it could provide a second, quasi-independent observable to test and calibrate 0$\nu\beta\beta$ decay NME calculations.  This is of special importance once data from experiments such as the Drift Chamber Beta-ray Analyzer \cite{Ish10} and Super-NEMO \cite{Piq06} become available.  

Finally, using the NEMO result \cite{NEMO08} for the 2$\nu\beta\beta$ ground-state transition rate, we find that the 2$\nu\beta\beta$ decay of $^{150}$Nd to the first excited 0$_1^+$ state in $^{150}$Sm is a factor of about 12 less likely than the decay to the 0$^+$ ground state. This observation is in contrast to what is expected for the 0$\nu\beta\beta$ decay of $^{154}$Sm \cite{Bel13}, for which the decay to the 0$_1^+$ state is favored over the ground-state decay.  It will be interesting to see the influence of the 1$^+_{sc}$ decay in $^{150}$Sm on the Interacting Boson Model 0$\nu\beta\beta$ and 2$\nu\beta\beta$ decay NME calculations for $^{150}$Nd \cite{Barea13}.

\begin{table*}
\caption[Summary of results.]{A summary of the results found in this work and compared to \cite{Bar09}. }
\begin{center}
\begin{tabular}{c|c|c|c|c|c} 
\hline
\hline
Transition                                      					& Gamma-Ray          & Efficiency		& $f_b$	& T$_{1/2}$ ($10^{20}$ y)							& T$_{1/2}$ ($10^{20}$ y)  \\
                                                					&  Energies (keV)	& (\%)      			&(\%)	&  (this work)			                 						  &  \cite{Bar09}\\
\hline
0$_1^+\rightarrow$2$_1^+\rightarrow$0$_{gs}^+$  &  333.97-406.52       	& 1.18$\pm$0.080  & 100	& $1.07^{+0.45}_{-0.25}(stat)\pm0.07(syst.))$			  	 & $1.33^{+0.36}_{-0.23}(stat.)^{+0.27}_{-0.13}(syst.)$ \\
2$_2^+\rightarrow$2$_1^+\rightarrow$0$_{gs}^+$  &  333.97-712.21       	& 0.316$\pm$0.021  &	89	& $>0.861$                              							   & $>8.0$ \\
2$_3^+\rightarrow$2$_1^+\rightarrow$0$_{gs}^+$  &  333.97-859.87       	& 0.316$\pm$0.021  &	45	& $>1.37$                                							 & $>5.4$ \\
0$_3^+\rightarrow$2$_1^+\rightarrow$0$_{gs}^+$  &  333.97-921.2       	& 0.708$\pm$0.040  &	91	& $>1.83$                                							 & $>4.7$ \\
\hline
\hline
\end{tabular}
\end{center}
\label{tab:summary}
\end{table*}

\section{Acknowledgements}
This work was supported in part by the US Department of Energy, Office of Nuclear Physics under grant No. DE-FG02-97ER41033.  The authors would like to thank R.B. Vogelaar for his role in establishing and maintaining KURF.  The authors would also like to thank A.S. Barabash for many helpful discussions and A.C. Crowell and B. Carlin for their networking expertise.

\end{document}